\documentclass[vecphys]{svmult}

\usepackage{makeidx}         % allows index generation
\usepackage{graphicx}        % standard LaTeX graphics tool
\usepackage{multicol}        % used for the two-column index
\usepackage[bottom]{footmisc}% places footnotes at page bottom
\usepackage{bm}% bold math
\def\journal#1#2#3#4{\ {#1}{\bf #2} ({#3})\  {#4}}
\def\NPB{\journal{Nucl.\ Phys.\ {\bf B}}}

\def\PRD{\journal{Phys.\ Rev.\ {\bf D}}}

\newenvironment{petitchar}{\begin{list}{}
{\leftmargin1.5em\rightmargin0.0cm}%
\item\small}{\end{list}}

\begin{document}

\title*{Non Perturbative Renormalization Group and Bose-Einstein Condensation}
\author{Jean-Paul Blaizot\inst{1}}
\institute{ECT*, Strada delle Tabarelle 286, I-38050 Villazzano (Trento), Italy
\texttt{blaizot@ect.it}}
\maketitle

%------------------------------------------------------------------------------

\def\bfphi{\mbox{\boldmath$\phi$}}
\def\bfvarphi{\mbox{\boldmath$\varphi$}}
\def\bfgamma{\mbox{\boldmath$\gamma$}}
\def\bfalpha{\mbox{\boldmath$\alpha$}}
\def\bftau{\mbox{\boldmath$\tau$}}
\def\bfnabla{\mbox{\boldmath$\nabla$}}
\def\bfsigma{\mbox{\boldmath$\sigma$}}
\def\bfpi{\mbox{\boldmath$\pi$}}

\def\0{\over } \def\2{{1\over2}} \def\4{{1\over4}}
\def\5{\hat } \def\6{\partial }

\def\a{\alpha } \def\b{\beta } \def\g{\gamma } \def\d{\delta }
\def\e{\epsilon } \def\z{\zeta } \def\h{\eta } \def\th{\theta }
\def\dh{\vartheta } \def\ka{\kappa } \def\l{\lambda }
\def\r{\rho } \def\s{\sigma }
\def\t{\tau } \def\o{\nix }
\def\ph{\varphi } \def\Ph{\phi } \def\PH{\Phi } \def\Ps{\Psi }
\def\O{\nix } \def\S{\nix } 
\def\Th{\Theta }
\def\L{\Lambda } \def\G{\Gamma } \def\D{\Delta }
\def\mn{{\mu\nu}}
\def\nbfgrad{\mbox{\boldmath$\nabla$}}

\newcommand \beq{\begin{eqnarray}}
\newcommand \eeq{\end{eqnarray}}

\newcommand \ga{\raisebox{-.5ex}{$\stackrel{>}{\sim}$}}
\newcommand \la{\raisebox{-.5ex}{$\stackrel{<}{\sim}$}}

\def\psib{\psi}
\def\phib{\phi}
\def\r{{\rm r}}
\def\d{{\rm d}}

\def \e {\mbox{e}}

\input epsf

% ---------- Gradient, etc.

\def\square{\hbox{{$\sqcup$}\llap{$\sqcap$}}}
\def\grad{\nabla}
\def\del{\partial}
% ---------- Fractions.

\def\frac#1#2{{#1 \over #2}}
\def\smallfrac#1#2{{\scriptstyle {#1 \over #2}}}
\def\half{\ifinner {\scriptstyle {1 \over 2}}
   \else {1 \over 2} \fi}

% ---------- Bras and kets

\def\bra#1{\langle#1\vert}
\def\ket#1{\vert#1\rangle}

%       \simge and \simle make the "greater than about" and the "less
% than about" symbols with spacing as relations.

\def\simge{\mathrel{%
   \rlap{\raise 0.511ex \hbox{$>$}}{\lower 0.511ex \hbox{$\sim$}}}}
\def\simle{\mathrel{
   \rlap{\raise 0.511ex \hbox{$<$}}{\lower 0.511ex \hbox{$\sim$}}}}

%       \buildchar makes a compound symbol, placing #2 above #1 and #3
% below it with \limits. \overcirc is a special case.

\def\buildchar#1#2#3{{\null\!
   \mathop#1\limits^{#2}_{#3}
   \!\null}}
\def\overcirc#1{\buildchar{#1}{\circ}{}}

%  \slashchar puts a slash through a character to represent contraction
%  with Dirac matrices. Use \not instead for negation of relations, and use
%  \hbar for hbar.

\def\slashchar#1{\setbox0=\hbox{$#1$}
   \dimen0=\wd0
   \setbox1=\hbox{/} \dimen1=\wd1
   \ifdim\dimen0>\dimen1
      \rlap{\hbox to \dimen0{\hfil/\hfil}}
      #1
   \else
      \rlap{\hbox to \dimen1{\hfil$#1$\hfil}}
      /
   \fi}

% ---------- Functions -- all defined like \sin, etc. in Plain TeX:

\def\real{\mathop{\rm Re}\nolimits}     % Re for real part
\def\imag{\mathop{\rm Im}\nolimits}     % Im for imaginary part

\def\tr{\mathop{\rm tr}\nolimits}       % tr for trace
\def\Tr{\mathop{\rm Tr}\nolimits}       % Tr for functional trace
\def\Det{\mathop{\rm Det}\nolimits}     % Det for functional determinant

\def\mod{\mathop{\rm mod}\nolimits}     % mod for modulo
\def\wrt{\mathop{\rm wrt}\nolimits}     % wrt for with respect to

% ---------- Abbreviations for units

\def\TeV{{\rm TeV}}                     % 10^12 electron volts
\def\GeV{{\rm GeV}}                     % 10^9  electron volts
\def\MeV{{\rm MeV}}                     % 10^6  electron volts
\def\KeV{{\rm KeV}}                     % 10^3  electron volts
\def\eV{{\rm eV}}                       % 1     electron volt

\def\mb{{\rm mb}}                       % 10^-27 cm^2
\def\mub{\hbox{$\mu$b}}                 % 10^-30 cm^2
\def\nb{{\rm nb}}                       % 10^-33 cm^2
\def\pb{{\rm pb}}                       % 10^-36 cm^2

%
% ----------Pictures macros-------
%

\def\picture #1 by #2 (#3){
  \vbox to #2{
    \hrule width #1 height 0pt depth 0pt
    \vfill
    \special{picture #3} % this is the low-level interface
    }
  }

\def\scaledpicture #1 by #2 (#3 scaled #4){{
  \dimen0=#1 \dimen1=#2
  \divide\dimen0 by 1000 \multiply\dimen0 by #4
  \divide\dimen1 by 1000 \multiply\dimen1 by #4
  \picture \dimen0 by \dimen1 (#3 scaled #4)}
  }

\def\centerpicture #1 by #2 (#3 scaled #4){
   \dimen0=#1 \dimen1=#2
    \divide\dimen0 by 1000 \multiply\dimen0 by #4
    \divide\dimen1 by 1000 \multiply\dimen1 by #4
         \noindent
         \vbox{
            \hspace*{\fill}
            \picture \dimen0 by \dimen1 (#3 scaled #4)
            \hspace*{\fill}
            \vfill}}

%---------------------------------------------------------

\def\figfermass{\centerpicture 122.4mm by 32.46mm
 (fermass scaled 750)}

%
%----------------------------------------------------------------------------

\section{Introduction}

These lectures are centered around a specific problem, the effect of weak repulsive interactions on the transition temperature $T_c$ of a Bose gas. This problem provides indeed a beautiful illustration of many of the techniques which have been discussed at this school on effective theories and renormalization group. Effective theories  are used first in order to obtain a simple hamiltonian  describing the atomic interactions: because the typical atomic interaction potentials are short range, and the systems that we consider are dilute, these potentials can be replaced by a contact interaction  whose strength is determined by the $s$-wave scattering length. Effective theories are used next in order to obtain a simple formula for the shift in $T_c$: one exploits there the fact that near $T_c$ the physics is dominated by  low momentum modes whose dynamics is most economically  described  in terms of classical fields; the ingredients needed to calculate the shift of $T_c$ can be obtained from this classical field theory. Finally the renormalization group is used both to obtain a qualitative understanding, and also as a non perturbative tool to evaluate quantitatively the shift in $T_c$.

In the  first lecture, I recall some   known aspects of Bose-Einstein condensation of the ideal gas. Then I turn to an elementary discussion of the interaction effects, and introduce an effective theory  with a contact interaction tuned to reproduce  the scattering length of the atom-atom interaction. I show that at the mean field level, weak repulsive interactions produce no shift in $T_c$. Finally, I briefly explain why approaching the transition from the low temperature phase is delicate and may lead to erroneous conclusions. 

In the second lecture I establish the general formula  for the shift of $T_c$:
\beq\label{deltaTc}
\frac{\Delta T_c}{T_c^0}=\frac{T_c-T_c^0}{T_c^0}=c(an^{1/3}), 
\eeq
where  $a$ is the s-wave scattering length for the atom-atom interaction, and $T_c^0$ the transition temperature of the ideal gas at density $n$. This formula holds in leading order in the parameter $an^{1/3}$ which measures the diluteness of the system.  Getting formula (\ref{deltaTc}) involves a number of steps. First I explain why perturbation theory cannot be used to calculate $\Delta T_c$, however small  $a$ is. Then I show that the problem with the perturbative expansion is localized in a particular subset of Feynman diagrams that are conveniently resummed by an effective theory of a classical 3-dimensional field.  The outcome of the analysis is the formula (\ref{deltaTc}) where $c$ is given by the following integral 
\beq\label{integralc0}
c\propto \int\frac{d^3 p}{(2\pi)^3}\left(\frac{1}{p^2+\Sigma(p)}-\frac{1}{p^2}\right).
\eeq
where the proportionality coefficient, not written here, is a known numerical factor, and $\Sigma$ is the self-energy of the classical field, whose calculation requires non perturbative techniques. 

In the last lecture, I use the non perturbative renormalization group (NPRG) in order to estimate  $c$. This  requires the knowledge of the 2-point function of the effective 3-dimensional field theory for all momenta, and in particular in the cross-over  between the  critical region of low momenta and  perturbative region of high momenta. This cross-over region is where the dominant contribution to the integral (\ref{integralc0}) comes from. In order to obtain an accurate determination of $\Sigma(p)$,  it has been necessary to develop new techniques to solve the NPRG equations.  Describing those techniques in detail would take us too far. I shall only present in this lecture the material that can help the student not familiar with the NPRG to understand how it works, and how it can be used. I shall do so by discussing several simple cases that at the same time provide indications on the approximation scheme that has been  developed in order to calculate $\Sigma(p)$. I shall end by reporting and discussing the results obtained with the NPRG, and compare them to those obtained using other non perturbative techniques. 

Recent discussion of Bose-Einstein condensation can be found in \cite{Pitaevskii03,Pethick02,RMP,Leggett:2001,Andersen:2003qj}. The equation (\ref{deltaTc}) for the shift of $T_c$ is derived and  discussed in the series of papers \cite{3/2club, BigN, bigbec, NEW,Fuchs04}. Much of the material of the last lecture is borrowed from the papers \cite{Blaizot:2004qa,Blaizot:2005wd,Blaizot:2006vr, Blaizot:2005xy, Blaizot:2006ak}.

\section{LECTURE 1.  Bose-Einstein condensation}

\subsection{ Bose-Einstein condensation for the non interacting gas}
\begin{itemize}
\item 
 The discussion of Bose-Einstein condensation of the ideal Bose gas in the grand canonical ensemble is standard. 
   We consider a homogeneous system of identical spinless bosons of mass $m$, at temperature $T$ . 
The occupation factor of a single particle state with momentum ${\bf p}$ is
\beq\label{occupquantique}
n_{\bf p}= \frac{1}{  {\rm e}^{(\varepsilon^0_{\bf p}-\mu)/T}-1},\qquad \varepsilon^0_{\bf p}=\frac{p^2}{2m},
\eeq
where $\mu$ is the chemical potential. The number density of   non-condensed 
particles is  given by
\beq\label{ndemuetT}
n=\int\frac{d^3{\bf p}}{(2\pi)^3}\, n_{\bf p}\equiv n(\mu, T).
\eeq

For small density, the chemical potential is negative and large in absolute value, ${\rm e}^{-\mu/T}\gg 1$. The gas is then described by Boltzmann statistics:
\beq\label{occupBoltz}
n_{\bf p}\approx {\rm e}^{-(\varepsilon^0_{\bf p}-\mu)/T},\qquad n\approx {\rm e}^{\mu/T} \lambda^{-3} ,
\eeq
where $\lambda$ is the thermal wavelength:
\beq\label{thermalwavel}
\lambda=\sqrt{\frac{2\pi}{mT}}.
\eeq
Unless specified otherwise, we use units such that $\hbar=1, k_B=1$. Boltzmann's statistics applies as long as $n\lambda^3\ll 1$, that is, as long as the  thermal wavelength is small compared to the interparticle distance. As one lowers the temperature, keeping the density fixed,   the  chemical potential increases, and so does the thermal wavelength. Eventually, as $\mu\to 0_-$, 
the density of non condensed particles reaches a maximum
\beq\label{thermaldensity}
n=\int\frac{d^3p}{(2\pi)^3} \frac{1}{  {\rm e}^{\varepsilon^0_{\bf p}/T}-1}=n(\mu=0,T)=\frac{\zeta(3/2)}{\lambda^3},
\eeq
where $\zeta(z)$ is the Riemann zeta-function and $\lambda$ the thermal wavelength.
As one keeps lowering the temperature, particles start to accumulate in the lowest energy single particle state. This is the onset of Bose-Eisntein condensation, which takes place then
 when
\beq\label{criticalline}
n\lambda^3= \zeta(3/2)\approx 2.612.
\eeq
  At this point the thermal wavelength has become comparable to the interparticle spacing. Eq.~(\ref{criticalline})  defines the critical line in the $n,T$ plane (see Fig.~\ref{fig1}). In particular, the critical temperature is given by
\beq
T_c^0= \frac{2\pi}{m}\left(\frac{n}{\zeta(3/2)}\right)^{2/3}.
\eeq
It is a function only of the mass of the atom, and of the density. 
\begin{figure}[ht]
  %\centering
  \includegraphics[width=12cm]{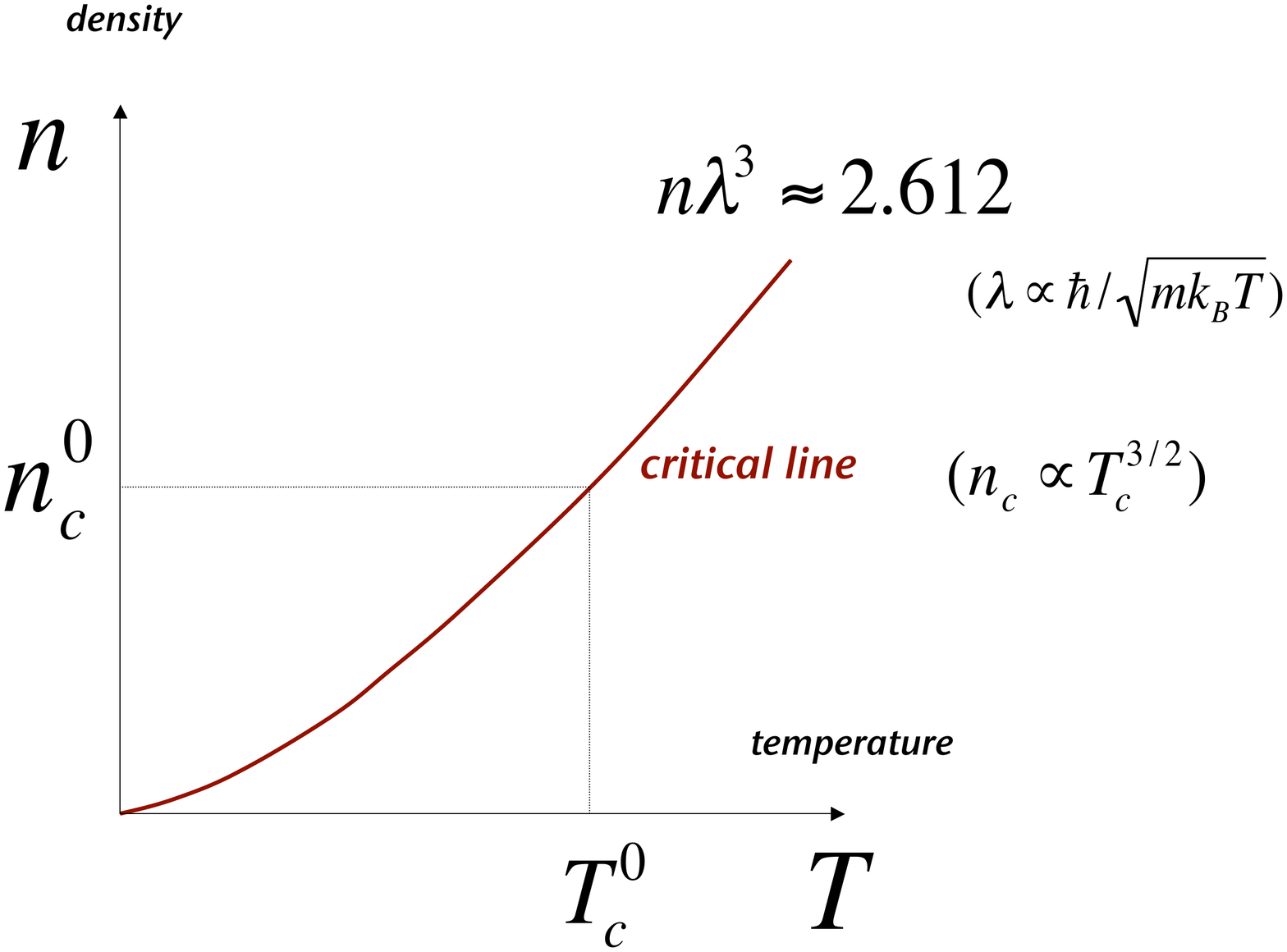}
  \caption{Critical line in the $(\mu,T)$ plane for the ideal Bose gas}
\label{fig1}
\end{figure}
For $T<T_c^0$, the chemical potential stays equal to zero and the single particle state ${\bf p}=0$ is macroscopically occupied, with density $n_0\propto n$ given by
\beq\label{condensatefraction}
n_0(T)=n\left(1-\left(\frac{T}{T_c^0}\right)^{3/2}\right).
\eeq

\item The occurrence of condensation relies on the existence of a maximum of the integral (\ref{ndemuetT})  fixing the number of non condensed atoms. This depends on the number of spatial dimensions. In  2 dimensions, 
 the integral   diverges logarithmically as $\mu\to 0_-$:
 \beq
n\lambda^2=-\ln\left(1-{\rm e}^{\mu/T} \right).
\eeq
In this case, there is no limit to the number of thermal particles, hence no condensation in the lowest energy mode. 

\item The Bose-Einstein condensation of the  non-interacting gas has some pathological features. In particular the compressibility  \beq
\frac{1}{n^2} \frac{\partial n}{\partial \mu}=\frac{1}{n^2}\frac{1}{T}\int\frac{d^3p}{(2\pi)^3}
n_{\bf p}(1+n_{\bf p}),
\eeq
diverges when $\mu\to 0_-$. Also the fluctutation of the number of particles in the condensate,   
\beq\label{fluctuationsN0}
\sqrt{\Delta\langle N^2\rangle}=\sqrt{N_0(N_0+1)}\simeq N_0,
\eeq
is of the same order of magnitude as the average value $N_0$. Such features can be related to the large degeneracy of states in Fock space that appears  as $\mu\to 0$. 

Consider indeed the ground state in Fock space of the hamiltonian of the non interacting Bose gas, 
\beq\label{H0}
H_0-\mu N= \sum_{\bf p}\left(\varepsilon^0_{\bf p}-\mu\right) a^\dagger_{\bf p} a_{\bf p}.
\eeq
If $\mu<0$, then the ground state is the vacuum, with no particle present in the system: adding a particle costs a positive ``free-energy" $-\mu$. When $\mu=0$, there appears a huge degeneracy  in Fock space: all the states with an arbitrary number of particles in the single particle state ${\bf p}=0$ are degenerate. If $\mu>0$, then there is no minimum, the more particle one adds in the state ${\bf p}=0$, the more free energy one gains ($-\mu$ for each added particle). Note that here the chemical potential does not fully plays its role of controlling the density:  either it is negative, and the density is zero, or it is positive and the density is infinite; if it vanishes the density is arbitrary. 

To relate the large degeneracy to the large fluctuations (\ref{fluctuationsN0}), one may use a simple maximum entropy technique to determine the most likely state with $N_0$ particles in the degenerate space. This calculation parallels the corresponding calculation done in the grand canonical ensemble at finite temperature. Maximizing
\beq
\hat S=-\sum_n p_n\ln p_n +\mu \sum_n n p_n-\lambda\sum_n p_n,
\eeq
with the constraints
\beq
\sum_n p_n=1,\qquad \sum_n np_n=N_0,
\eeq
one finds
\beq
p_n=\frac{1}{Z}{\rm e}^{\mu n},\qquad Z=\frac{1}{1-{\rm e}^{\mu }},\qquad \mu=-\ln\left(1+\frac{1}{N_0}\right).
\eeq
For large $N_0$,
\beq
p_n\approx \frac{{\rm e}^{-n/N_0}}{N_0}, 
\eeq
which leads indeed to the large fluctuations (\ref{fluctuationsN0}). 

Of course, some of the pathologies of the ideal Bose gas in the grand canonical ensemble could be eliminated by working in the canonical ensemble, where the particle number is fixed (see e.g. \cite{Ziff77}). However, we shall keep the discussion within the grand canonical ensemble, as it is closer to familiar field theoretical techniques. This allows us in particular to treat Bose-Eisntein condensation (of the interacting gas) as a symmetry breaking phenomenon. 

The large degeneracy of states in Fock space implies that infinitesimal interactions could have a large effect, and indeed they do: as soon as weak repulsive interactions are present the  large fluctuations are damped, and the compressibility becomes finite .

\end{itemize}

\subsection{Interactions in  the dilute gas}
\begin{itemize}
\item As we shall discuss later in this lecture, the dominant effect of the interactions in the dilute gas   can be accounted for by an effective contact interaction whose strength $g$ is proportional to the $s$-wave scattering length $a$:
\beq\label{aetg}
V({\bf r_1-r_2})=g \delta ({\bf r_1-r_2}),\qquad  g=\frac{4\pi a}{m}.
\eeq

\item  In the mean field (Hartree-Fock) approximation, which is also the leading order in $a$,   the effect of the interaction is a simple shift of the single particle energies:
\beq\label{HFspectrum}
\varepsilon_{\bf p}^0\longrightarrow\varepsilon_{\bf p}^0+2gn,
\eeq
where the factor 2 comes from the exchange term.

 It is easy to see that this produces no shift in $T_c$: because the shift of the single particle energy is constant, independent of the momentum, eq.~(\ref{ndemuetT}) which gives the number density of non condensed particles, can be written
\begin{equation}
n=n (\mu -\Delta \mu,T)  \label{ig3}
\end{equation}
with
\begin{equation}
 \Delta \mu = 2g n   \label{ig4}.
\end{equation}
Bose-Einstein condensation now occurs when $\mu-\Delta
\mu=0$, instead of $\mu=0$ in the non interacting case, but clearly  the critical line is identical to that given by eq.~(\ref{criticalline}).

 Of course, because the Hartree-Fock self-energy in eq.~(\ref{HFspectrum}) depends on the density, the relation
between the chemical potential at the transition and the critical density is
more complicated than in the non-interacting case.  The transition now takes place at a finite value of the chemical potential, and  some of the pathologies of the ideal gas disappear. 
In particular the compressibility remains finite at $T_c$ (fluctuations remain important though):
\beq
\frac{\partial n}{\partial \mu}= \frac{\frac{1}{T}\int_p n_p(1+n_p)}{1+\frac{2g}{T}\int_p n_p(1+n_p)}\simeq\frac{1}{2g}.
\eeq

\item The presence of the interactions allows us to treat the phenomenon of Bose Einstein condensation as a symmetry breaking phenomenon. Let us return first to the ground state, and calculate the  free energy \beq\label{freeenergy0}
\frac{\Omega(n_0)}{{\cal V}}= -\mu n_0+\frac{g}{2} n_0^2,
\eeq
where ${\cal V}$ is the volume. $\Omega(n_0)$ is the expectation value of the hamiltonian (\ref{H0}) to which is added the interaction (\ref{aetg}), $V\sim (g/2)  a^\dagger_0 a^\dagger_0 a_0 a_0$, in a coherent state  containing an average number of particles  $N_0=n_0{\cal V}$ in the state ${\bf p=0}$, $\ket{n_0}\sim \exp{\sqrt{N_0}a^\dagger_0}\ket{0}$. 
In contrast to what happened for the ideal gas, it is now possible to minimize $ \Omega(n_0)$  w.r.t. $n_0$ in order to find the optimum ground state for a given $\mu$. One gets
\beq
n_0=\mu/g.
\eeq
Thus the ground state has now a finite number of particles, and the fluctuations are normal, proportional to the square root of the mean value. The density in the ground state is $n=\partial \Omega/\partial \mu=n_0$, and the pressure is $P=-\Omega(n_0=\mu/g)/{\cal V}=gn^2/2$, so that also at zero temperature the compressibility is  finite, $\chi^{-1}=ndP/dn=gn^2$. Note however that the coherent state is a state with a non definite number of particles: the symmetry related to particle number conservation is spontaneously  broken. 
 We shall return later to the similar picture at finite temperature, and  come back to this issue of symmetry breaking.

\end{itemize}

\subsection{Atoms in a trap}

Although our main discussion  concerns  homogeneous systems, it is instructive to contrast the situation in  homogeneous systems to what happens   in a   trap. We shall consider here a spherical harmonic trap, corresponding to the following external potential
\beq
V(r)=\frac{1}{2}m\omega_0^2 r^2.
\eeq

\begin{itemize}
\item Consider first the non interacting gas. We  assume the validity of a semiclassical approximation allowing us to express the particle density as the following phase space integral:  
\beq\label{densitysemiclas}
n(r)=\int\frac{d^3p}{(2\pi)^3}\frac{1}{{\rm e}^{(\varepsilon(r,p)-\mu)/T}-1},
\eeq
where $\varepsilon(r,p)=p^2/2m+(1/2)m\omega_0^2 r^2$. This requires that the temperature is large compared to the level spacing, $k_BT\gg \hbar\omega_0$, a condition well satisfied near the transition if $N$, the number of particles in the trap, is large enough. The number density in the trap can be written as 
\beq
N=\int d^3{\bf r}\, n(\mu-\half m\omega_0^2 r^2,T),
\eeq
where $n(\mu,T)$ is the function (\ref{ndemuetT}). Thus in a wide trap for which the semiclassical approximation is valid, the particles experience the same conditions as in a uniform system with a local  effective chemical potential $\mu- \half m\omega_0^2 r^2$. In particular, 
eq.~(\ref{densitysemiclas}) shows that the density at the center of the trap is related to the chemical potential $\mu$ by the same relation as in the homogeneous gas. It follows that $T_c^0$ is determined by the same condition as for the homogeneous gas, that is, $n(0)\lambda^3=2.612$ where $n(0)$ is the density at the center of the trap. An explicit calculation gives 
\beq\label{Tc0intrap}
\frac{k_BT_c^0}{\hbar\omega_0}=\left(\frac{N}{\zeta(3)}\right)^{1/3},
\eeq
with $\zeta(3)=1.202$. Note that while the condensation condition is ``universal" when expressed in terms of the density at the center of the trap, the dependence of $T_c^0$ on $N$  depends on the form of the confining potential. In the present case, the $N$ dependence can be obtained from the following heuristic argument. For a temperature $T\simge T_c^0$ the particle density is approximately given by the classical formula
\beq\label{napproxtrap}
n\sim {\rm e}^{-{m\omega_0^2 r^2}/{2T}},
\eeq
so that the thermal particles occupy a cloud of radius $R_{th}\sim \sqrt{T/m\omega_0^2}\sim a^2_{ho}/\lambda$, where  $\a_{ho}=1/\sqrt{m\omega_0}$ is the characteristic radius of the harmonic trap and $\lambda\sim 1/\sqrt{mT}$ is the thermal wavelength. The average density in the thermal cloud is $\bar n\sim N R_{th}^{-3}$. At the transition, the interparticle distance is of the order of the thermal wavelength, so that, $\bar n^{1/3}\sim N^{1/3}\lambda/a^2_{ho}\sim \lambda^{-1}$, or $N^{1/3} \lambda^2\sim a^2_{ho}$, from which the relation $T_c^0\sim \omega N^{1/3}$ follows.

Note that the confining potential makes condensation easier than in the uniform case. This is related to  the fact that the density of single particle states in a trap decreases more rapidly with decreasing energy  than in a uniform system: it goes as $\varepsilon^{d-1}$ in a trap and as $\varepsilon^{d/2-1}$ in a uniform system, where $d$ is the number of spatial dimensions.  It follows in particular that, in a trap, condensation can occur in $d=2$, in contrast to the homogeneous case;  the heuristic argument presented above yields then $T_c^0\sim \omega N^{1/2}$. Note however that the effects of the interactions in a 2-dimensional trap are  subtle (for a recent discussion see \cite{PNAS}).

\item The leading effect of repulsive interactions in a trap is to push the particles away from the center of the trap,  thereby decreasing the central density. This effect is analogous to that produced by an increase of the temperature. One expects therefore the interactions to lead to a decrease of the transition temperature. To estimate this, we note that, in the presence of interactions,  the density is still given by eq.~(\ref{napproxtrap}) after substituting $(1/2)m\omega_0^2 r^2\to (1/2)m\omega_0^2 r^2+2gn(r).$ A simple calculation gives then the density $n(r)$ (for $r$ not too large, and to leading order in $a$) in terms of the density $n(0)$ at the center of the trap \cite{Fuchs04}:
\beq
n({\bf r})&\approx& n(0) {\rm e}^{-\beta (m\omega_0^2 r^2/2) (1-2gn(0)/T)}
\nonumber\\
&\approx& n(0)\left[ 1 -\frac{1}{2T}\frac{m\omega_0^2r^2}{1+4a\lambda^2 n(0)} \right] .
\eeq
This result suggests that the effect of the interaction can also be viewed as a modification of the oscillator frequency, $\omega_0^2\to \omega_0^2/(1+4a\lambda^2n(0)$. This is enough to estimate the shift in $T_c$: \beq
\frac{\Delta T_c}{T_c^0}=\frac{\Delta\omega}{\omega_0}\sim - a\lambda^2 n(0)\sim -\frac{a}{a_{ho}} N^{1/6},
\eeq
where, in the last step,  we have used the fact that $n(0)\sim \lambda^{-3}$ at the transition. A more  explicit calculation yields the proportionality coefficient 
 $-1.32$ \cite{Giorgini96}.

It is important to keep in mind that this effect of mean field interactions on $T_c$ is very different from the one that leads to eq.~(\ref{deltaTc}) (indeed the sign of the effect is different). If we were comparing systems at fixed central density rather than at fixed particle number, there  would be no shift of $T_c$ (see the discussion in \cite{Fuchs04}).

\end{itemize}

\subsection{The two-body problem}

We now come back to the  construction of the effective interaction that can be used in many-body calculations of the dilute gas. We shall in particular recall how  effective field theory can be used to relate the effective interaction to the low energy scattering data. More on the use of effective theories can be found in the lecture by T. Sch\"{a}fer in this volume \cite{Schafer:2006yf}. A  pedagogical introduction is given in Ref.~\cite{Kaplan:2005es}.

Consider two atoms of mass $m$ interacting with the central 
two-body potential $V(r)$, with ${\bf r}$ the relative coordinate
and $r=|{\bf r}|$. The relative wave function satisfies the
Schr\"odinger equation \beq
\left[-\frac{\hbar^2\,\nabla^2}{m}+ V(r)\right]\psi({\bf
r})=E\psi({\bf r}). \eeq The scattering wave function is given by
($E=\hbar^2 k^2/m$) \beq \psi_{\bf k}({\bf r})={\rm e}^{i{\bf
k}\cdot {\bf r}}+f({\bf k}',{\bf k}) \frac{{\rm e}^{ikr}}{r}, \eeq
where  ${\bf k}$ is the initial relative momentum (with $k=|{\bf
k}|$), while ${\bf k}'$ is the final relative momentum.  The
scattering is elastic and because the potential is  rotationally invariant,
 the scattering amplitude $f({\bf k}',{\bf k})$ is   a
function only of  the scattering angle   $\theta$  in the center of
mass frame, and of the energy $E=\hbar^2k^2/m$.  For short-range interactions, the interaction takes place predominantly in the $s$-wave, and the scattering amplitude becomes of a function of the energy only. It has then the following low momentum expansion: \beq\label{ascattampl}
f(E)=\frac{1}{-\frac{1}{a}+\frac{r_0}{2}k^2-ik} \qquad \hbar k=\sqrt{mE}.
\eeq 
where $a$ is the scattering length, $r_0$ the effective range, and the neglected terms in the denominator involve higher powers of $k$.
In the very low momentum limit, when $kr_0\,ka\ll1$, one can ignore the effective range. Then the scattering amplitude depends on  a single parameter, $a$.  

The scattering amplitude can be expressed as the matrtix element between plane wave states of the T-matrix: 
\beq\label{defT}
\langle {\bf k'} |T|{\bf k}\rangle=-\frac{4\pi\hbar^2}{m} f({\bf k',k}),
\eeq
and the T-matrix itself can be calculated in terms of the Green function $G_0$
\beq
T=V-VG_0T,
\eeq
where
\beq
G_0(E)=\frac{1}{H_0-E},
\eeq
and $E\to E+i\eta$ with $E$ real  for the retarded Green's function $G_0^{R}$. 
The following formal relations are useful
\beq
H=H_0+V, \qquad G^{-1}=H-E=G_0^{-1}+V,
\eeq
and
\beq
T=V-VGV=V-VG_0T\qquad T^{-1}=V^{-1}+G_0.
\eeq
Note that both $G(E)$ and $T(E)$ are analytic functions of $E$, with poles on the negative real axis corresponding to the energies of the bound states, and a cut on the real positive axis.

Let us now turn to the many-body problem. Assuming that the atoms interact only through the two body potential $V(r)$, we can write the interaction hamiltonian as
\beq\label{Vpotential}
V=\frac{1}{2}\int d^3 {\bf r_1}\d^3{\bf r_2}\, \psi^\dagger({\bf r_1})\psi^\dagger({\bf r_2})V({\bf r_1-r_2})\psi({\bf r_2})\psi({\bf r_1}),
\eeq
We demand that the local effective theory
\beq
V_{eff}=\frac{g_0}{2}\int d^3 {\bf r}\, \psi^\dagger({\bf r})\psi^\dagger({\bf r})\psi({\bf r})\psi({\bf r}),
\eeq
reproduce the same scattering data in the two particle channel at low momentum as the original potential $V({\bf r_1-r_2})$. We know that in the long wavelength limit the scattering amplitude depends only on the scattering length $a$, so we expect $g_0$ to be related to $a$. 

To establish this relation we calculate the scattering amplitude for the effective theory. For a contact interaction $T(E)$ is given by \beq
\frac{1}{T(E)}=\frac{1}{g_0}+G_0^R(E), \eeq 
where
\beq\label{Gretardeddelta}
G_0^{R}(E)=\int\frac{d^3p}{(2\pi)^3}\frac{1 }{p^2/m-E-i\eta}
\eeq
To calculate
$G_0^R(E)$, which is divergent, we introduce a cut-off
$\Lambda$ on the momentum integral \beq
G^{R}(E)&=&m\int\frac{d^3p}{(2\pi)^3}\frac{1 }{p^2-Em}=\frac{m}{2\pi^2}\int_0^\Lambda dp \frac{p^2}{p^2-mE-i\eta}\nonumber\\
&=&\frac{m\Lambda}{2\pi^2}
+\frac{m^2E}{4\pi^2}\int_{-\infty}^{+\infty}dp\frac{1}{p^2-mE-i\eta} .\eeq
It is then convenient to define a ``renormalized" strength $g_R$ by
\beq
\label{Vrenorm}\frac{1}{g_R}=\frac{1}{g_0}+\frac{m\Lambda}{2\pi^2},\qquad
g_R=\frac{4\pi a}{m}, \eeq where in the last relation $a$ is the
scattering length. 
This relation between $g_R$ and $a$ is obtained by comparing $T(E)$ obtained from eq.~(\ref{Gretardeddelta}) ($E>0)$  \beq
 \frac{1}{T(E)}=\frac{1}{g_R}+i\frac{m}{4\pi}\sqrt{mE},
\eeq
 with eq.~(\ref{ascattampl}), and using the relation (\ref{defT}) to relate $T(E)$ and $f(E)$.
\begin{petitchar}
\renewcommand{\baselinestretch}{.90}\small
\noindent{\bf Remark.} 
One may improve the description by including the effective range correction. This is done by adding to the hamiltonian a term of the form \cite{Braaten:2000eh}
\beq
\frac{g_0'}{2}\int d^3 r\, \bfnabla(\psi^\dagger\psi) \cdot \bfnabla(\psi^\dagger\psi),
\eeq
and adjusting $g_0'$ so as to reproduce the scattering amplitude of the original two body problem, at the precision of the effective range. At tree level in the effective theory,  the calculation of the scattering amplitude yields
\beq
\bra{{\bf k_3 k_4}}T\ket{{\bf k_1\,k_2}}&=& -2g_0'\left[ ({\bf k_1-k_3})\cdot ({\bf k_2-k_4})+({\bf k_1-k_4})\cdot ({\bf k_2-k_3})\right]\nonumber\\
&=&8g_0' k^2,
\eeq
where in the second line $k$ is the magnitude of the relative momentum. Note that in order to make the identification with the two-body problem discussed above, we have to pay attention that the two-body problem traditionally treats the two particles as distinguishable, whereas the present calculation involves matrix element of the $T$-matrix between symmetric two particle states. Thus the tree level calculation of the $T$-matrix for the hamiltonian (\ref{Vpotential}) yields $\bra{{\bf k_3 k_4}}T\ket{{\bf k_1\,k_2}}=2g_0$
which differs by a factor 2 with the conventional definition of the scattering length. Staying with the usual convention, we therefore write 
\beq
T\approx \frac{4\pi}{m}\frac{a}{1-r_0ak^2/2+ika}\approx \frac{a}{1+ika}\left(1+\frac{ar_0}{2}k^2\right),
\eeq
from which the identification of $g_0'$ follows
\beq
g_0'=g_0\frac{ar_0}{8}.
\eeq
\end{petitchar}
For recent discussions on the application of effective field theory techniques to the Bose gas, see \cite{Braaten:1996rq,Braaten:1997ky,Braaten:2000eh}.

\subsection{One-loop calculation}

Having at our disposal an effective many-body hamiltonian, we may now perform detailed calculations of the effect of the interactions. The one loop calculation that 
we present here   gives us the opportunity to come back to the issue of symmetry breaking, illustrates the use of the delta potential and points to the difficulty of approaching the phase transition from below. 

The grand canonical partition function can be written as  a path integral:
\begin{eqnarray}\label{ZBose}
Z&=&{\rm Tr}{\rm e}^{-\beta \hat H-\mu N}=\int _{\psi(\beta)=\psi(0)}{\cal D}(\psi,\psi^{*}) \;{\rm e}^{-{\cal S}}\;, \eeq
with 
\beq \label{BoseAction}{\cal S}\!=\! \int_0^{\beta} d\tau \int {d^3 {\bf r}} \left\{
 \psi^* (x) \left(\frac{\partial}{\partial \tau}\!-\! \frac{\nabla^2}{2m}\,
\!-\!\mu \right) \psi (x)+\frac{g_0}{2}
\psi^*(x)
\psi^* (x)
\psi(x)\psi(x)
\right\},\nonumber\\
\end{eqnarray}
and $\beta=1/T$ is the inverse temperature. The complex field $\psi(x)=\psi(\tau, {\bf r})$ to be integrated over is a periodic function of the imaginary time $\tau$, with period $\beta$. The action ${\cal S}$ is invariant under a $U(1)$ symmetry:
\beq\label{U1symmetry}
\psi(x)\rightarrow {\rm e}^{i\alpha }\psi(x),\qquad \psi(x)^*\rightarrow {\rm e}^{-i\alpha }\psi^*(x).
\eeq

It is convenient to add an external source $j(x)$ linearly coupled to the
bosonic field (thereby breaking the $U(1)$ symmetry): 
\beq
Z[j]=\int_{\psi(\beta)=\psi(0)}{\cal D}(\psi^*,\psi)\,{\rm e}^{-{\cal S}_j},
\eeq
where
\beq
{\cal S}\rightarrow {\cal S}={\cal S}-\int {\rm d}^3 x \left[j(x)\psi^* (x)+j^*(x)\psi(x)\right].
\eeq

The loop expansion is an
expansion around the field configurations which make the action stationnary.  These field configurations, which we denote by  $\phi_0(x)$ are
determined by the  equation
\beq\label{stationary}
\left[\frac{\partial}{\partial \tau} - \frac{\hbar^2}{2m}\nabla^2 -\mu\right ] \phi_0(x) +g_0|\phi_0|^2
\phi_0(x)=j(x)
\eeq
and its complex conjugate.  For uniform,
time-independent, external sources
$j$, eq.~(\ref{stationary}) admits constant field solutions (compare   eq.~(\ref{freeenergy0})):
\beq
\left[-\mu+g_0|\phi_0|^2\right]\phi_0=j.
\eeq
There are two solutions. For $j=0$, $\phi_0=0$ or $|\phi_0|^2=\mu/g_0$. The solution $\phi_0=0$ corresponds to the vacuum state and is the stable solution when $\mu<0$. For $\mu>0$ (and $g_0>0$) the
second solution corresponds to broken $U(1)$ symmetry and BE condensation; in that case the solution $\phi_0=0$ is a
maximum of the free energy.

We now expand the field around the solution $\phi_0$ which corresponds to a minimum of the (classical) free energy: 
\beq
\psi(x)\longrightarrow \phi_0+\psi(x)\qquad \psi(x)^*\longrightarrow \phi_0^*+\psi(x)^*\, ,
\eeq
with $\psi(x)$ having only $k\ne 0$ Fourier components, 
and keep in ${\cal S}$ terms which are at most quadratic in the fluctuations. That is, we write ${\cal S}_j={\cal S}_0+{\cal S}_1$, with ${\cal S}_0$ the
classical action
\beq
{\cal S}_0=\beta V \left[ -\mu|\phi_0|^2+\frac{g_0}{2} \left( |\phi_0|^2 \right)^2 +j\phi_0^*+j^*\phi_0\right],
\eeq
and $S_1$ the one-loop correction
\beq
{\cal S}_1&=&\int_0^\beta {\rm d}\tau \int {\rm d}^3 x 
\left[ \psi^*(x) \left(   \frac{\partial}{\partial \tau} - \frac{\hbar^2}{2m}\nabla^2 -\mu+2g_0|\phi_0|^2\right) \psi(x)  \right.\nonumber\\
&&\left.
+ \frac{g_0}{2}\,\left( (\phi_0^*)^2\,\psi(x)\psi(x)+(\phi_0)^2\,\psi^\dagger(x)\psi^\dagger(x)\right)
\right].
\eeq
The gaussian integral is standard  and, after a Legendre transform to eliminate $j$, it yields the following  expression   of the thermodynamic potential:
\beq\label{ToyodaOmega}
\frac{\Omega}{{\cal V}}=\frac{g_R n_0^2}{2}\!-\!\mu n_0\!+\!\frac{1}{2{\cal V}}\sum_{k\ne
0}(E_k-\varepsilon_k)-\frac{g_R^2n_0^2}{4{\cal V}}\sum_k
\frac{1}{\varepsilon_k^0}+\frac{1}{\beta {\cal V}}\sum_k\ln\left( 1-{\rm e}^{-\beta E_k}
\right)\nonumber\\
\eeq
where $E_k$ is the Bogoliubov quasiparticle energy:\beq
E_k^2=(\varepsilon_k^0+2gn_0-\mu)^2-(gn_0)^2.
\eeq
Note that in this calculation we have used the relation (\ref{Vrenorm}) in order to replace the bare coupling constant by the renormalized one. This is the origin of the fourth term in the r.h.s. of eq.~(\ref{ToyodaOmega}) which eliminates the divergence in the sum over the zero point energies. Note that this replacement assumes that $g_R$ and $g_0$ are perturbatively related, which can only be possible if the cut-off $\Lambda$ in (\ref{Vrenorm})  is not too large, i.e., if $m\Lambda g_R/(2\pi^2)\ll 1$, or $\Lambda\ll \pi/(2a)$.

This calculation was first made in the context of Bose-Einstein condensation by Toyoda \cite{Toyoda}, and led him to the erroneous conclusion that $\Delta T_c\sim -\sqrt{an^{1/3}}$. It is essentially a  mean field calculation, the mean field being entirely due to the  condensed particles.  This approximation, equivalent to the lowest order Bogoliubov theory (see e.g. \cite{FW}), 
describes correctly the ground state at zero temperature and
its elementary excitations. However, its extension near the critical temperature meets
several difficulties; in particular it predicts a first order phase transition
\cite{BG}, a point apparently overlooked in Ref.~\cite{Toyoda}.  In fact, above $T_c$,  $\phi_0=0$, and the present one-loop calculation  yields  the free energy  of an ideal gas, with no shift in
the critical temperature.

%%%%%%%%%%%%%%%%%%%%%%%%%%%%%%%%%%%%%%%%%%%%%%%%%%
%%%%%%%%%%%%%%%%%%%%%%%%%%%%%%%%%%%%%%%%%%%%%%%%%%

\section{LECTURE 2.  The formula for $\Delta T_c$}

\begin{figure}[ht]
  %\centering
  \includegraphics[width=12cm]{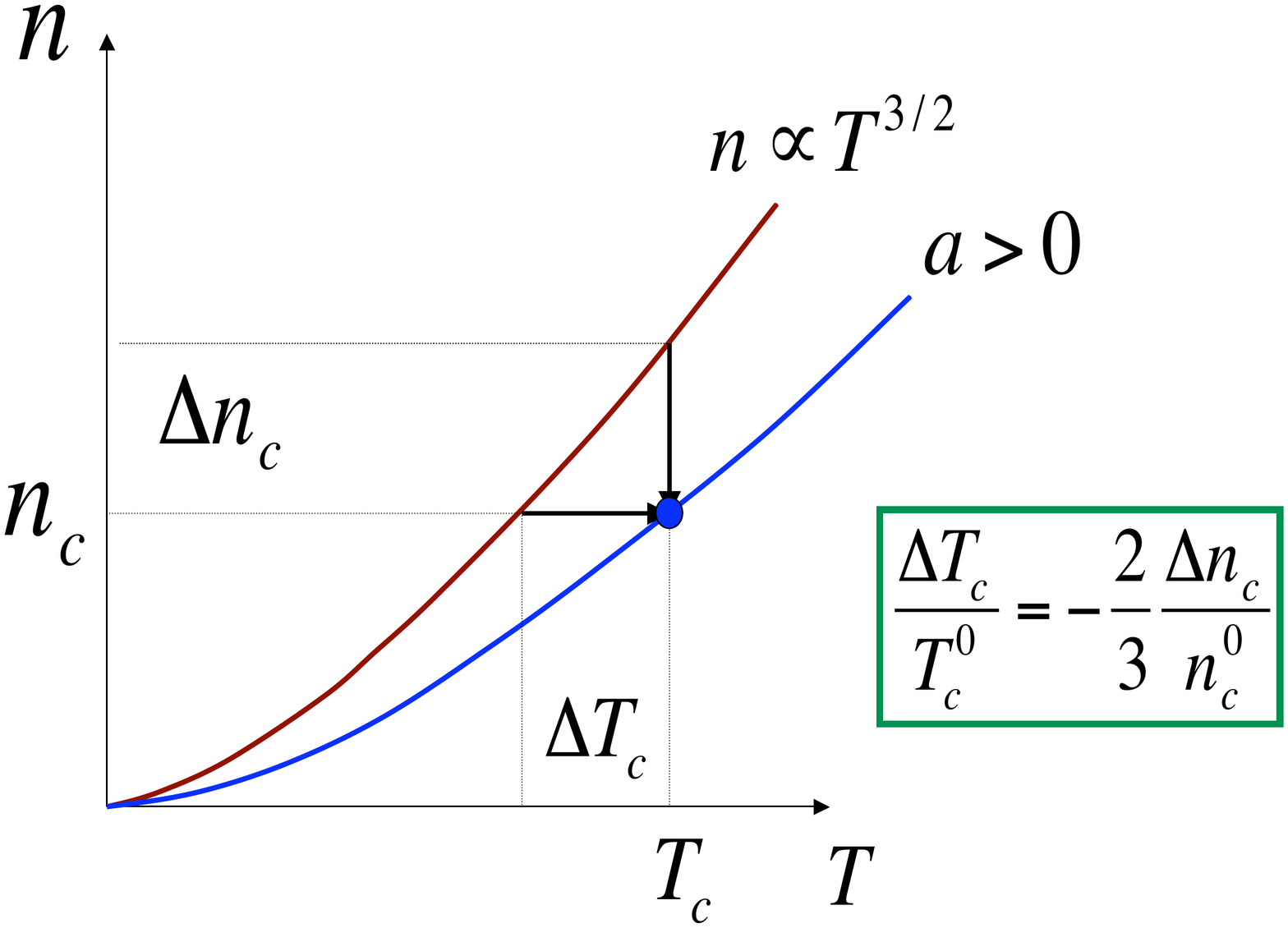}
  \caption{Expected change in the critical line caused by weak repulsive interactions characterized by a small scattering length $a$
}
\label{fig3}
\end{figure}

 Let us start by considering again the phase diagram in Fig. 1. We want to determine the change of the critical line due to weak repulsive interactions. For small and positive values of the $s$-wave scattering length $a$,  we expect the change illsutrated in Fig.~\ref{fig3}, with the two critical lines close to each other.  One can then relate, in leading order, the shift of the critical temperature at fixed density, $\Delta T_c$, to that of the critical density at fixed temperature, $\Delta n_c$. Since, for $a=0$,  $n_c^0\propto (T_c^0)^{3/2}$, we have
\beq\label{deltaTdeltan}
\frac{\Delta T_c}{T_c^0}=-\frac{2}{3}\frac{\Delta n_c}{n_c^0}.
\eeq
It turns out that it is easier to calculate at fixed temperature than at fixed density, and in the following we shall set up the calculation of $\Delta n_c$.

In the previous lecture, we have seen that, under appropriate conditions, the effective hamiltonian is of the form
\beq \label{Hamiltonian}H=\int{d^3 {\bf r}} \left\{
 \psi^\dagger({\bf r}) \left(\-\frac{\bfnabla^2}{2m}\,
\!-\!\mu \right) \psi ({\bf r})+\frac{g_0}{2}
\psi^\dagger({\bf r})
\psi^\dagger ({\bf r})
\psi({\bf r})\psi({\bf r})
\right\},\nonumber\\
\end{eqnarray}
where $g_0$ is related to the scattering length by (see eq.~(\ref{Vrenorm})):
\beq
\label{Vrenorm2}\frac{1}{g_R}=\frac{1}{g_0}+\frac{m\Lambda}{2\pi^2},\qquad
g_R=\frac{4\pi a}{m}.\eeq
This effective hamiltonian provides an accurate description of phenomena where the dominant degrees of freedom have long wavelength, $kr_0\ll 1$, and the system  is dilute, $an^{1/3}\ll 1$. Recall also that we shall be working in the vicinity of the transition where  $n^{1/3}\lambda\simeq 1$, with $\lambda$ the thermal wavelength.

\subsection{Condensation condition and critical density}

In order to exploit standard techniques of quantum field theory and many-body physics (see e.g.\cite{FW,KB,self-consistent,BR}), we shall first relate the particle density to the single particle propagator. 
The particle density can be written as
\beq\label{netG}
n&=& \langle \psi^\dagger({\bf r})\psi({\bf r})\rangle, \nonumber\\
&=& \lim_{\tau\to 0_-}\langle{\rm T}  \psi(\tau,{\bf r})\psi^\dagger(0,{\bf r}) \rangle=  \lim_{\tau\to 0_-}  G(\tau,0),
\eeq   
where       
\beq
\psi^\dagger(\tau,{\bf r})={\rm e}^{\tau H}\psi^\dagger({\bf r}) {\rm e}^{-\tau H},\qquad\qquad
  \psi(\tau,{\bf r}) ={\rm e}^{\tau H}\psi({\bf r}) \, {\rm e}^{-\tau H},
\eeq
are respectively  the creation and annihilation operators in the Heisenberg
representation  and T denotes (imaginary) time ordering (see e.g. \cite{BR}). 
The single particle propagator is 
\beq
G(\tau_1-\tau_2, {\bf r_1-r_2})&=&\langle {\rm T} \psi(\tau_1,{\bf r_1}) \psi^\dagger (\tau_2,{\bf r_2})\rangle\nonumber\\ &=&{\rm Tr}\left(\frac{{\rm e}^{-\beta
H}}{Z} {\rm T} \psi(\tau_1,{\bf r_1}) \psi^\dagger (\tau_2,{\bf r_2})\right).
\eeq
It is a periodic function of  $\tau=\tau_1-\tau_2$: for  $0<\tau<\beta$,  $G(\tau-\beta,{\bf r})=G(\tau,{\bf r})$, where $\beta=1/T$ is the inverse temperature. Because of its periodicity, it can be represented by a Fourier series
\beq
\label{G_de_tau}
 G(\tau,{\bf p})=T\sum_n
e^{-i\omega_n\tau}G_\alpha(i\omega_n,{\bf p}),
\eeq
where the $\omega_n$'s are called the Matsubara frequencies:
\beq
\omega_n = 2n\pi T, \eeq
and we have also taken a Fourier transform with respect to the spatial coordinates. We are  making here an abuse of notation: we denote by the same symbol the function and its Fourier transform, with the implicit understanding that the arguments, whether space-time or energy-momentum variables, are enough to specify which function  one is considering. 
The inverse transform is given by
\beq\label{G_de_omega}
G(i\omega_n,{\bf p}) = \int_0^\beta {\rm d}\tau\, e^{i\omega_n\tau}G(\tau,{\bf p}).\eeq

In the absence of interactions, the hamiltonian is of the form
\beq
H_0-\mu N&=& \int d^3{\bf r} \;\psi^\dagger({\bf r})\left( -\frac{\hbar^2\bfnabla^2}{2m}-\mu\right) \psi({\bf r})\nonumber\\ &=&\sum_{\bf p} \left(\varepsilon^0_{\bf p}-\mu\right) a^\dagger_{\bf p} a_{\bf p}=\sum_p \;\varepsilon_{\bf p}\, a^\dagger_{\bf p} a_{\bf p}, 
\eeq
where 
\beq
\psi({\bf r})=\sum_{\bf p}\,\frac{{\rm e}^{i{\bf p\cdot\bf r} }}{\sqrt{\cal V}}\,a_{\bf p},\qquad \psi^\dagger({\bf r})=\sum_{\bf p}\,\frac{{\rm e}^{-i{\bf p\cdot\bf r} }}{\sqrt{\cal V}}\,a^\dagger_{\bf p}, \qquad \varepsilon^0_{\bf p}=\frac{{\bf p}^2}{2m}.
\eeq
In these formulae,  ${\cal V}$ is the volume of the system, and  the creation and annihilation operators satisfy $[a_{\bf p},a^\dagger_{\bf p'}]=\delta_{\bf p,\bf p'}$. The free single particle propagator can  be obtained by a direct calculation. It reads
\beq\label{freeG0}
G^{-1}_0(i\omega_n,{\bf p})=\varepsilon_{\bf p}-i\omega_n,
\eeq 
or, in imaginary time, 
\beq\label{freeG}
G_0(\tau_1-\tau_2,{\bf p}) &=& \langle{\rm
T}a_{\bf p}(\tau_1)a_{\bf p}^\dagger(\tau_2)\rangle_0\nonumber\\
&=& e^{-\varepsilon_{\bf p}(\tau_1-\tau_2)}
\left[(1+
n_{\bf p})\theta(\tau_1-\tau_2)+ n_{\bf p}\theta(\tau_2-\tau_1)\right], 
\eeq
where:
\beq
n_{\bf p}\equiv \langle a^\dagger_{\bf p}
a_{\bf p}\rangle_0= \frac{ {\rm Tr}{\rm e}^{-\beta(H_0-\mu N)}a^\dagger_{\bf p} a_{\bf p}}{{\rm Tr}{\rm e}^{-\beta(H_0-\mu N)}}=\frac{1}{e^{\beta\varepsilon_{\bf p}}- 1},\qquad \varepsilon_{\bf p}=\varepsilon_{\bf p}^0-\mu.
\eeq
Thus, for non interacting particles, 
\beq
\lim_{\tau\to 0_-}  G_0(\tau,{\bf p})=n_{\bf p},
\eeq
so that the formula (\ref{netG}) yields the familiar formula (\ref{ndemuetT}) of the density.

The full propagator is related to the bare propagator by Dyson's equation
\beq\label{GSigma}
G^{-1}(i\omega_n,{\bf p})=G_0^{-1}(i\omega_n,{\bf p})+\Sigma(i\omega_n,{\bf p}),
\eeq
where $\Sigma(i\omega,{\bf p})$ is the self-energy.
In this case eq.~(\ref{netG}) yields the following expression for the density
\beq\label{density2}
n= \lim_{\tau\to 0_-} T\sum_n\int\frac{d^3 {\bf p}}{(2\pi)^3}\frac{ e^{-i\omega_n\tau}}{\varepsilon_{\bf p}-i\omega_n+\Sigma(i\omega_n,{\bf p})}.
\eeq 
or equivalently  the occupation factor in the interacting system            
\beq\label{occupfactor2}
n_{\bf p} =
\lim_{\tau\to 0_-} T\sum_n e^{-i\omega_n\tau}G(i\omega_n,{\bf p}).
\eeq

We shall approach the condensation from the high temperature phase. Then the system remains in the normal state all the way down to $T_c$. 
        The Bose-Einstein  condensation occurs when the chemical potential reaches a value such that (see e.g. \cite{PPbook}):
\begin{equation}
G^{-1}(\omega=0,{\bf p}=0)=0 \quad {\rm or} \quad \Sigma
(\omega=0,{\bf p}=0)=\mu \,.  \label{g3}
\end{equation}
        At that point,
\begin{equation}
G^{-1}(i\omega_n,{\bf p})=i\omega_n-\varepsilon_{\bf p}^0-\left[\Sigma(i\omega_n,{\bf p})
     -\Sigma(0,0)\right].  \label{g4}
\end{equation}

By using the general relation (\ref{density2}) between the Green function and the density, one can then write the following formula for the shift $\Delta n_c$ in the critical density caused by the interaction: 
\beq
\Delta n_c&=& \lim_{\tau\to 0_-} T\sum_n e^{-i\omega_n\tau}\!\int\frac{d^3 {\bf p}}{(2\pi)^3} \nonumber\\&& \times \left\{    \frac{1}{\varepsilon_{\bf p}^0+\Sigma(i\omega_n,{\bf p})-\Sigma(0,0)-i\omega_n}- \frac{1}{\varepsilon_{\bf p}^0-i\omega_n} 
\right\}.
\eeq
The first term in this expression is the critical density of the interacting system at temperature $T$, the second term is the critical density of the non interacting system at the same temperature. This formula makes it obvious that $\Delta n_c$ vanishes if the self-energy is independent of energy and  momentum. This is the case in particular when the  interactions are treated at the mean field level,  as we already observed. In this case, $\Sigma=2gn$, and the formula above yields
\beq
\Delta n_c=\int\frac{d^3 {\bf p}}{(2\pi)^3}\left\{   \frac{1}{{\rm e}^{\beta(\varepsilon_{\bf p}^0+2gn-\mu_c)}-1}-\frac{1}{{\rm e}^{\beta \varepsilon_{\bf p}^0}-1}  \right\}=0,
\eeq
which vanishes since $\mu_c=2gn$. 
\begin{figure}
\begin{center}
\includegraphics[height=3cm]{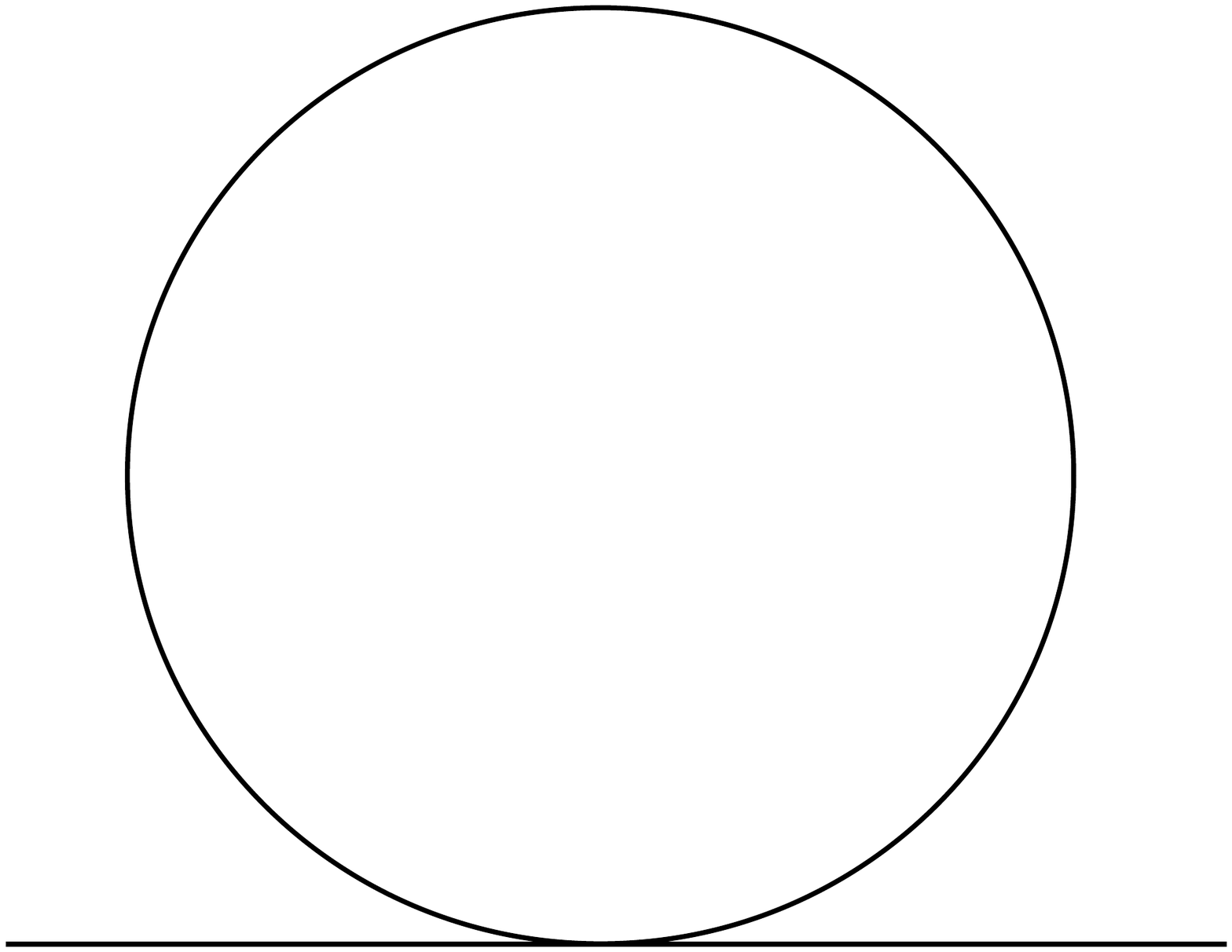}
\end{center}
\caption{\label{fig:tadpole}
The contribution to the self-energy $\Sigma$ that is of leading order in $a$}
\end{figure}

\subsection{Breakdown of perturbation theory}

 Because the interactions are weak, one may imagine calculating  $\Delta n_c$ by perturbation theory.  However the
perturbative expansion for a critical theory does not exist for any fixed
dimension $d<4$; infrared divergences prevent a complete calculation, as we shall recall.   If one
introduces an infrared cutoff $k_c$ to regulate the momentum integrals, one
finds that perturbation theory breaks down when $k_c \sim a/\lambda^2$, all
terms being then of the same order of magnitude.  

The leading order on $a$ is given by the diagram in Fig.~\ref{fig:tadpole}. As we just saw, the contribution of this diagram to $\Sigma$ is just the mean field value $2gn$, and the net effect on $\Delta n_c$ is zero. We shall then examine  the second order contribution, given by the diagram displayed in Fig.~\ref{figsunset}. We shall  see that this diagram is infrared divergent. Next, we shall show, using simple power counting, that such infrared divergences occur in higher orders and signal a breakdown of perturbation theory as one approaches the critical point.

\subsubsection{Second order perturbation theory}

\begin{figure}
\begin{center}
\includegraphics[height=5cm]{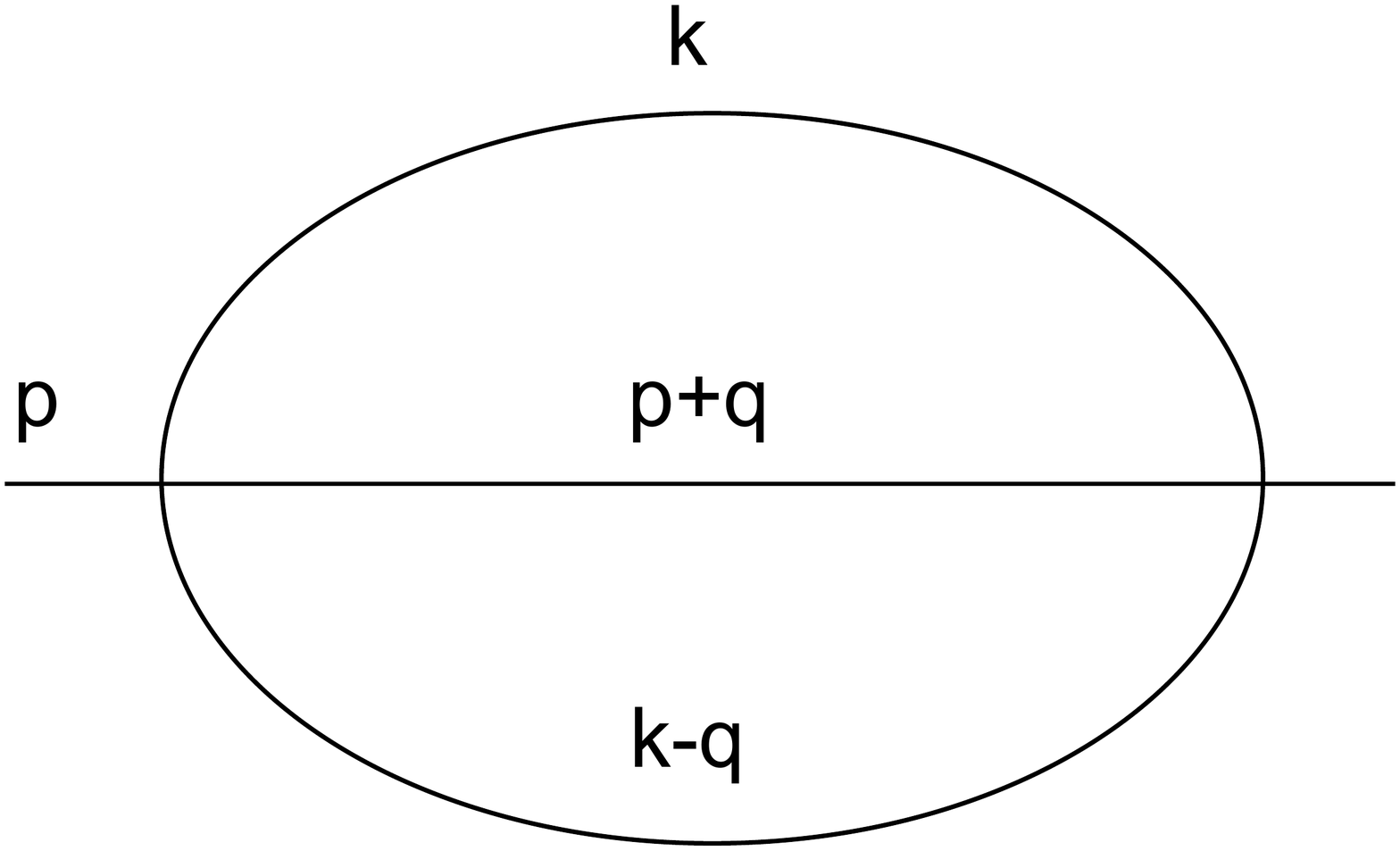}
\end{center}
\caption{\label{figsunset}
The contribution to the self-energy $\Sigma$ that is, a priori,  of second order in $a$}
\end{figure}

The second order
self-energy diagram  is the lowest order diagram that is momentum dependent  and can therefore yield corrections to the critical density. It is displayed in Fig.~\ref{figsunset}.
Its contribution is given by 
\beq
\Sigma(i\omega_n,{\bf p})&=&-2g^2 T^2 \sum_{n' n''}\int \frac{d^3 {\bf k}}{(2\pi)^3}
\frac{d^3 {\bf q}}{(2\pi)^3}\nonumber \\
&\times& \frac{1}{\varepsilon_{\bf k-q}-i(\omega_{n'}-\omega_{n''})} 
\,\frac{1}{\varepsilon_{\bf k}-i\omega_{n'}} \,\frac{1}{\varepsilon_{\bf p+q}-i(\omega_{n}+\omega_{n''})}.\nonumber\\
\eeq
Anticipating that infrared divergence can occur, we focus on the contribution where $\omega_{n'}=\omega_{n''}=0$, and calculate the difference $\Sigma(i\omega_n=0,{\bf p})-\Sigma(i\omega_n=0,{\bf p}=0)$, which we write from now on simply as  $\Sigma_{cl}({\bf p})-\Sigma_{cl}(0)$:
\beq
\Sigma_{cl}({\bf p})-\Sigma_{cl}(0)&=&
-2 g^2 T^2  \int \frac{d^3 {\bf k}}{(2\pi)^3}
\frac{d^3 {\bf q}}{(2\pi)^3} \nonumber\\
&\times&\frac{1}{(\varepsilon^0_{\bf k-q}-\mu')(\varepsilon^0_{\bf k}
-\mu') }
\left[ \frac{1}{\varepsilon^0_{\bf p+q}-\mu')}
- \frac{1}{\varepsilon^0_{\bf q}-\mu'} \right].\nonumber\\
\label{sigsumnew}
\eeq

In this calculation we have used Hartree-Fock propagators, and set 
\beq\label{zeta}
\mu'=\mu-\Sigma_{HF}\equiv -\frac{k_c^2}{2m}, 
\eeq
where $\Sigma_{HF}=2gn$.
The quantity $k_c$, whose significance will appear more clearly as we progress in the discussion, plays the role of an infrared cutoff in the integrals. Note that $k_c \rightarrow 0 $ ($\mu ^{\prime }\rightarrow 0$) when
$T\rightarrow T_{c}^{0}$. With this new notation  \begin{equation}
\varepsilon^0_{k}-\mu ^{\prime }=(k^{2}+k_c^2)/2m.  \label{epsilon}
\end{equation}
We shall also set
\beq\label{defU}
U(p)\equiv 2m(\Sigma_{cl} ( p)-\Sigma_{cl} (0)).
\eeq

The integral in eq.~(\ref{sigsumnew}) can be calculated analytically (see \cite{bigbec}) and yields
\beq
U(p)=128\pi ^{2}\left( \frac{a}{\lambda ^{2}}\right)
^{2}\left\{ \frac{3k_c}{ p}{\arctan}\frac{p}{3k_c}+\frac{1}{2}\ln
\left[ 1+\left(\frac{p }{3k_c} \right)^{2} \right]-1 \right\} .  \label{finitepot}
\eeq
This equation shows that  $U(p)$ is a monotonically
increasing function of
$p$, $\sim p^2$ at small $p$, and  growing logarithmically at large $p$.
This logarithmic behavior, obtained in perturbation theory, remains
in general the
dominant behavior of
$U(p)$ at large $p$,
i.e., for
$k_c\ll   p \, \la \, 1/\lambda$. This equation (\ref{finitepot}) also reveals the anticipated infrared divergence: $U(p)$ diverges logarithmically as $k_c\to 0$. 

 The condensation condition  (\ref{g3}), $\mu'=\Sigma_{cl}(0)$, reads
 \beq
k_c^2\approx  128\pi ^{2}\left( \frac{a}{\lambda
^{2}}\right)
^{2}\left[ \ln \left(\frac{\Lambda   }{3k_c}\right)-\gamma \right],  \label{sigmadiv}
\eeq
where the right hand side is $-2m\Sigma_{cl} (0)$, $\gamma = 0.577 \ldots$ is Euler's constant, and the approximate equality is valid when
$\Lambda\gg
k_c$, which we assume to be  the case. Here  $\Lambda$ is an  ultraviolet cut-off that we have introduced to handle the divergence of $\Sigma_{cl} (0)$, and whose  origin is  the neglect of the non vanishing Matsubara frequencies. Typically,  $\Lambda\sim 1/\lambda$. Eq.~(\ref{sigmadiv}) shows that  
$k_c\sim a/ \lambda^2$ when $a$ is small \cite{bigbec}. 

This simple calculation  illustrates the
limits of a pertubative approach. The infrared cutoff $k_c\sim a/\lambda^2$ introduces  spurious $a$ dependence. The
condensation condition which relates the infrared cutoff
to the  microscopic length $\lambda$, induces a spurious logarithmic
correction which does not vanish as
$a\to 0$. In fact such logarithms do appear as higher order corrections, as we shall see later, but are absent in the leading order result.

\subsubsection{Higher orders}

The infrared divergences that we have identified in the second order calculation persist, and worsen, in  higher order contributions. This may be seen by using a simple  power counting argument.  Let
us first consider
     diagrams in which all the internal lines carry vanishing  Matsubara
frequencies. We use again HF propagators so that all the
functions that are integrated in the diagrams  are products
of fractions of the
form
\begin{equation}
\left[ K^{2}+k_c^2\right] ^{-1},
\label{sum}
\end{equation}
where $K$ denotes a generic combination of momenta; it is then
natural to use the dimensionless products $K/k_c $ as new
integration variables.
Consider then a diagram of order $a^{n}$. The lowest order $n=2$ has
just been calculated, and, for large $p/k_c$,  it is proportional to $ (a/\lambda
)^{2}\ln(p/k_c)$, where $p$ is the external momentum.
For $n>2$, every additional order brings in one factor $a$ from the
vertex,  one
integration over three-momenta, a factor $T$, and two  internal
propagators.
The contribution of the diagram can thus be written as:
\begin{equation}\label{series}
T\,\left(\frac{a}{\lambda }\right)^{2}\left(\frac{a }{k_c \lambda
^{2}}\right)^{n-2}F(p/k_c ),
\label{cl8bis}
\end{equation}
where $F$ is a dimensionless function, which we do not explicitly need 
here. The main point is that when one approaches the
critical temperature, the
coherence length becomes large so that the summation of terms (\ref{series})
diverges. In the
critical region, $k_c \sim a/\lambda^2$, so that
all the terms in
the perturbative expansion are of the same order of magnitude. Therefore, at the
critical point, perturbation theory is not valid.

Let us now assume that in a given diagram some 
propagators carry
non-zero Matsubara frequencies so that one momentum integration ($k$) will
be altered. For that
integration, the presence of a non  vanishing Matsubara frequency in the denominators of the propagators ensures that no
singularity at $k=0$
can take place. Essentially, in the corresponding
propagators, $k_c$ is
replaced by a term proportional to $1/\lambda$, so that  one factor
$a/\lambda ^{2}k_c$ in (\ref{cl8bis}) is now replaced by $
a/\lambda $. Compared 
to the diagram with only vanishing Matsubara frequencies,
this diagram is down by
a factor $a/\lambda$, and thus negligible in a leading order calculation
of $\Sigma$. This reasoning generalizes trivially to diagrams containing more non vanishing Matsubara frequencies. 

\subsubsection{Formula for $\Delta n_c$}

It follows from the previous discussion that in order to obtain the leading order shift in the critical density,    one may retain in eq.~(\ref{density2}) the contribution of the zero Matsubara frequency only. That is,
\beq\label{deltanc2}
\Delta n_c&\simeq&T \int\frac{d^3 p}{(2\pi)^3} \left\{    \frac{1}{\varepsilon_p+\Sigma_{cl}(p)-\Sigma_{cl}(0)}- \frac{1}{\varepsilon_p}\right\} \nonumber\\
&\simeq& -\frac{2}{\pi\lambda^2}\int_0^\infty dp  \frac{U(p)}{p^2+U(p)},
\eeq
with
$U(p)$ given by eq.~(\ref{defU}). Note that this integral is finite: $U(p)\sim \ln p$ at large $p$, and $p^2+U(p)\sim p^{2-\eta}$ at small $p$. Note also that since $U(p)>0$ (in fact the general qualitative behavior of $U(p)$ is correctly given by eq.~(\ref{finitepot})), the correction $\Delta n_c$ is negative, implying a positive shift of $T_c$.

\subsection{Classical field approximation}

    Once restricted to their zero Matsubara frequency components, the fields
$\psi$ and $\psi^\dagger$ can be considered as classical fields, and the
entire calculation can be cast in terms of a classical field theory.  
To see that, let us expand the field
variables in the path integral (\ref{ZBose}) in terms of their Fourier
components:
\beq
\psi(\tau,{\bf r})=\psi_0({\bf r})+T \sum_{n\ne 0} {\rm e}^{-i\omega_n \tau}
\psi_n({\bf r}),
\eeq
where the $\omega_n$'s are the Matsubara frequencies, and we have called $\psi_0({\bf r})$ the $\tau$-independent part of the field. 

\begin{center}
\begin{figure}
\includegraphics[scale=.13,angle=0]{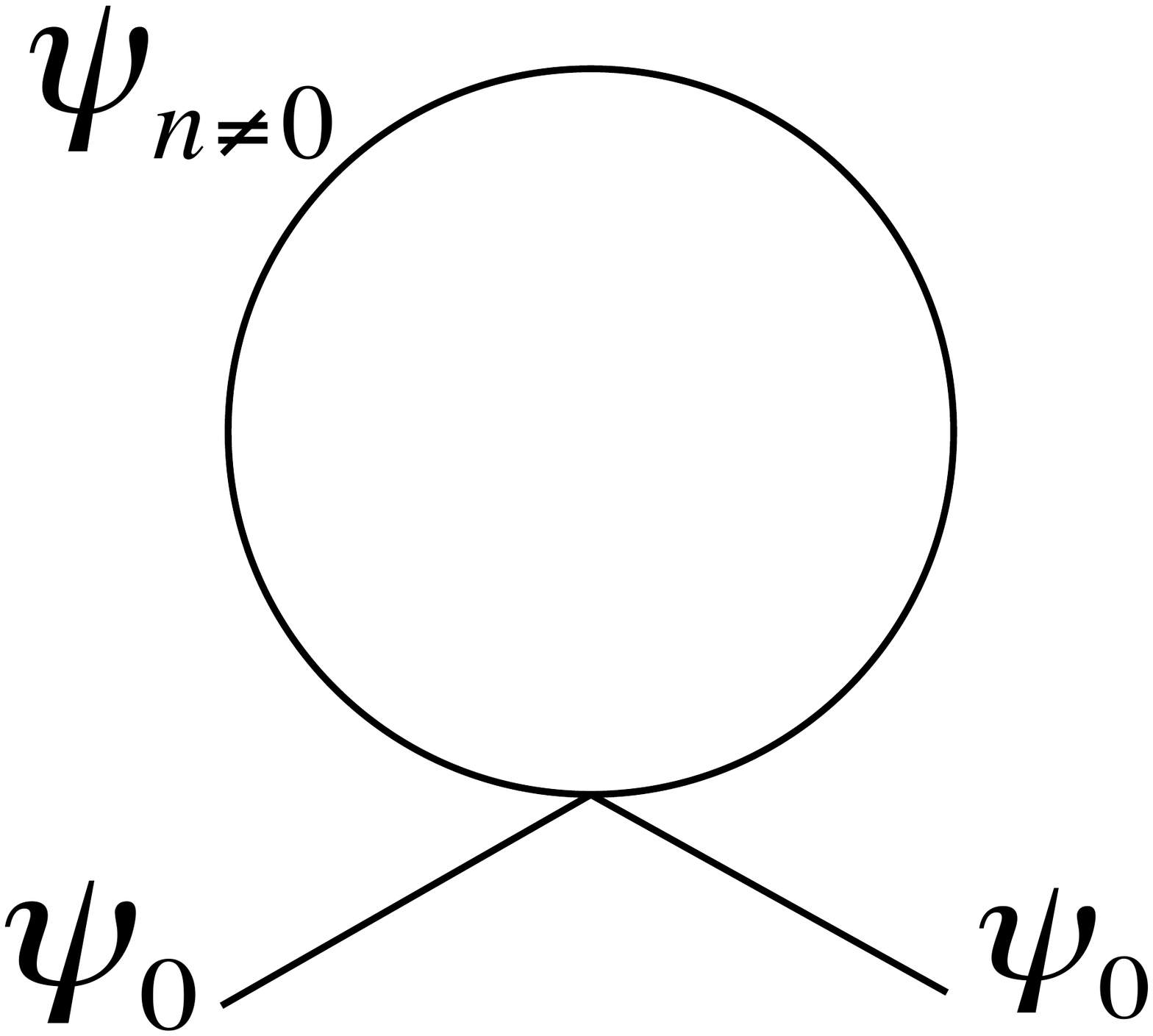}\includegraphics[scale=.13,angle=0]{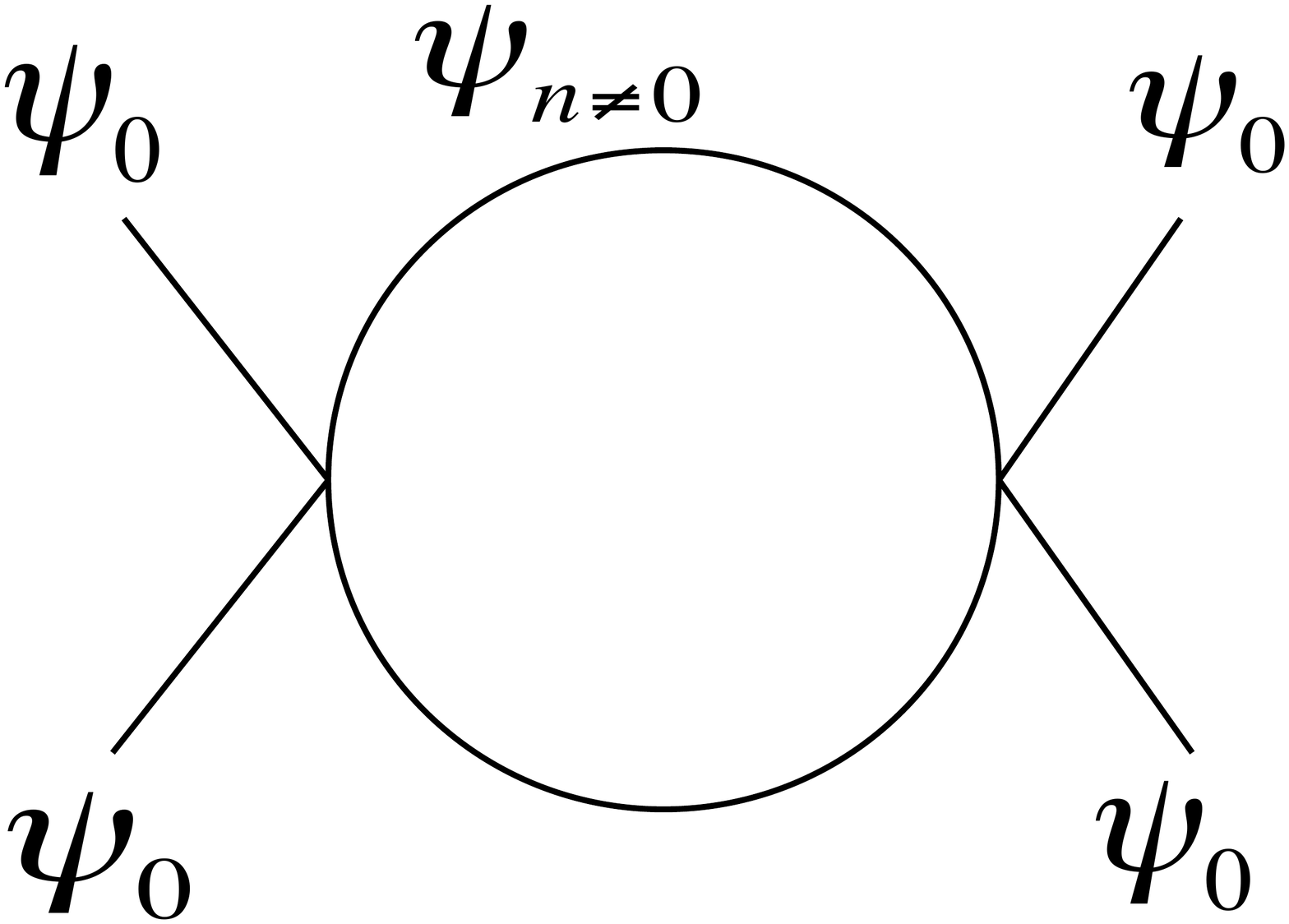}\includegraphics[scale=.13,angle=0]{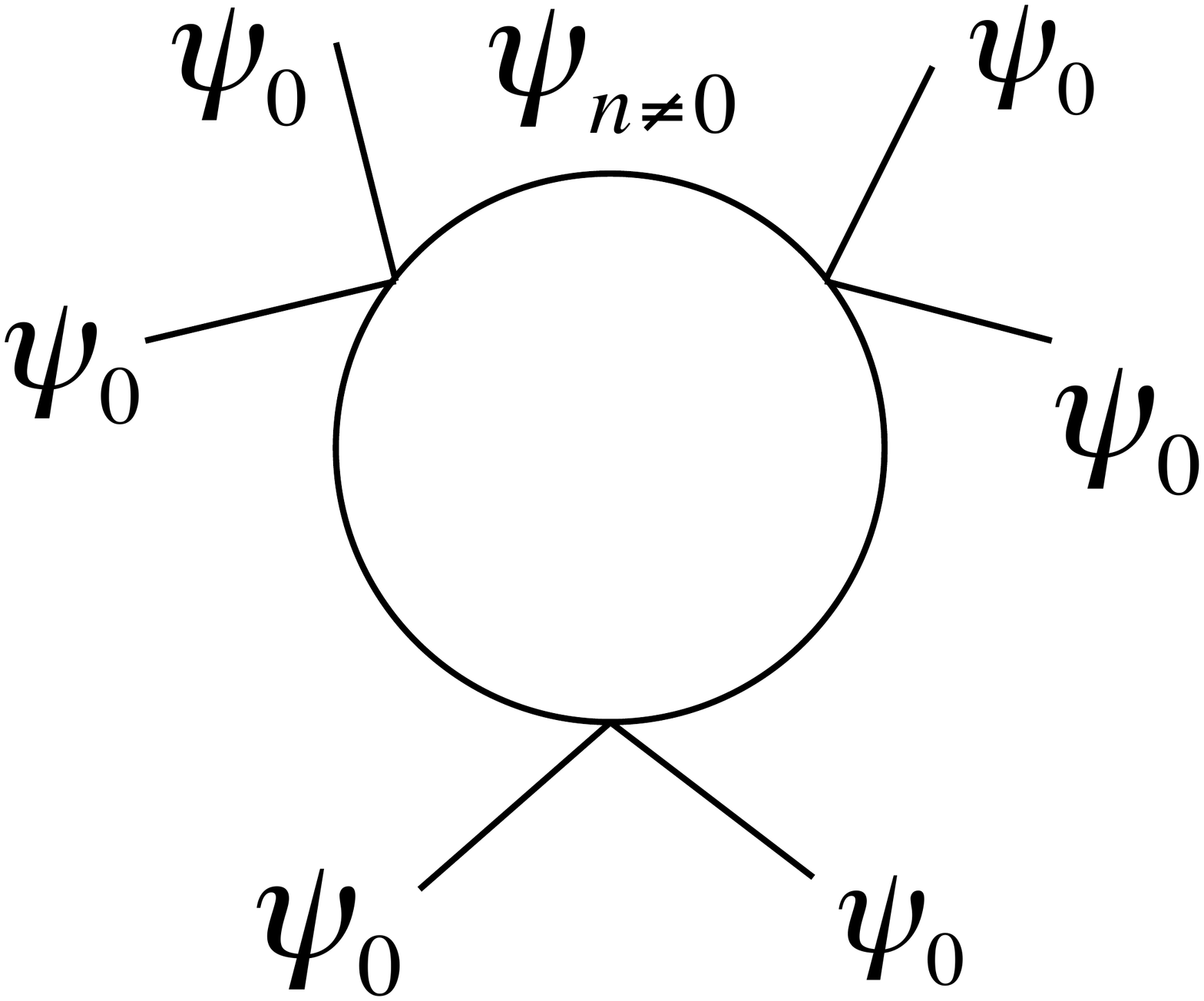}
\caption{\label{fig:effectivetheory}Some diagrams contributing to the parameters of the effective theory. The external legs represent the field $\psi_0$ carrying vanishing Matsubara frenqency. The internal lines carry non vanishing Matsubara frequencies $\omega_n$. The first diagram from the left is a correction to the chemical potential; it is of order $a$, but does not contribute to the shift in $T_c$. The second diagram is a correction to the $\psi_0^4$ coupling constant. The last diagram is the contribution to the $\psi^6$ vertex. The last two contributions are of order $a^2$ and $a^3$ respectively, and can be ignored in a leading order analysis.}
\end{figure}
\end{center}

The partition function  (\ref{ZBose}) can then be written as:
\beq\label{ZCE0}
Z\,=\,{\cal N}_1
\int{\cal D}\psi_0\exp\left\{-S_{\rm eff}[\psi_0]\right\}\,,\eeq
where $\psi_0$ depends only on spatial coordinates, and
\beq\label{ZCE1}
\exp\left\{-S_{\rm eff}[\psi_0]\right\}=\,{\cal N}_2
\int{\cal D}\psi_{n\ne 0}\,{\rm e}^{-{\cal S}}\,\eeq
where ${\cal S}$ is the action (\ref{BoseAction}). ${\cal N}_1$ and ${\cal N}_2$ are (infinite) normalization constants. 
The quantity $S_{\rm eff}[\psi_0]$ is the effective action for  $\psi_0$.   Aside from the 
direct classical field
contribution to which we shall return shortly,
this effective action  receives also contributions which,
diagrammatically, correspond to connected
diagrams whose external lines are associated to
$\psi_0$, and  the internal lines are the
propagators of the non-static
modes $\psi_{n}$. A few examples are displayed in  Fig.~\ref{fig:effectivetheory}. Thus, a priori,  $S_{\rm eff}[\psi_0]$ contains
operators of arbitrarily high order in $\psi_0$.
In the present case, however, it is easy to verify that the contributions beyond those kept in the classical action, are of higher order in $a$, and can therefore be ignored in a leading order calculation. 

The strategy which consists of integrating out the non-static modes
     in perturbation theory in order to obtain an effective
three-dimensional  theory for the soft static modes, is  
referred  to, in another context, as ``dimensional reduction'' (see e.g. 
\cite{Gins80,Appel81,Nadkarni83}). 

In leading order, the effective action is the restriction of eq.~(\ref{BoseAction}) to the 
static mode $\psi_0$  and the
partition function (\ref{ZCE0})  can be written as:
\beq\label{ZCEcl}
Z\,\approx\,{\cal N}
\int{\cal D}\psi_0\,\exp\left\{
- \beta \int{\rm d}^3r \,({\cal H}({\bf r})-\mu {\cal N}({\bf r}))\right\}\,,\eeq
where { $\psi_0({\bf r})$ is   a three-dimensional
field, and}
\beq\label{LEcl}
{\cal H-\mu N}\,=\,\frac{|\nbfgrad\psi_0|^2}{2m}-\mu' 
\,|\psi_0|^2 + \frac{g_0}{2}(|\psi_0|^2)^2\,.\eeq
We shall refer to this limit as the {\it classical field
approximation}\index{classical field approximation}. 
The zero Matsubara component of the density is given by
$\langle |\psi_0(r)|^2 \rangle$, which diverges in the effective theory. However, recall that by
assumption, the wavenumbers of the classical field are limited to
$k$ less than an ultraviolet cutoff
$\Lambda \sim \lambda^{-1}$. In fact we shall not need to use a cut-off since the variation of the critical density is a finite quantity (see eq.~(\ref{eONrho}) below).

\begin{petitchar}
\renewcommand{\baselinestretch}{.90}\small
\noindent{\bf Remark.} 

Ignoring the time dependence of the fields is equivalent to retaining only
the zero Matsubara frequency in their Fourier decomposition. Then the Fourier
transform of the free propagator is simply:
\beq
G_0({\bf p})\,=\,\frac{T}{\varepsilon_{\bf p}}\,.
\eeq
This may be obtained directly from  (\ref{G_de_tau}) and (\ref{freeG0}) keeping only the term
with $\omega_n=0$, or from eq.~(\ref{ZCEcl}). The classical field approximation corresponds also to the following approximation of the statistical factor (see eq.~(\ref{occupfactor2}))
\beq\label{BESOFT}
n_{\bf p}=\frac{1}{{\rm e}^{\beta\varepsilon_{\bf p}} -1}\approx
\frac{T}{\varepsilon_{\bf p}}\,.
\eeq
Both approximations make sense only for $\varepsilon_{\bf p}\ll T$, implying
$n_{\bf p}\gg 1$. In this limit, the energy per mode is
$\propto\varepsilon_{\bf p} n_{\bf p}\approx T$,  as expected from the
classical equipartition theorem. This long wavelength limit can also be viewed as a high temperature limit (the time dependence of the field becomes indeed unimportant as $\beta\to 0$). 

 One should not confuse this classical field approximation with the
classical limit reached when the thermal wavelength of the particles
becomes small compared to their average separation distances. In this
limit, the occupation of the single particle states becomes small, and
the statistical factors can be approximated by their Boltzmann form:
\beq
\frac{1}{{\rm e}^{\beta(\varepsilon_p-\mu)}\pm 1}\approx {\rm
e}^{\beta(\varepsilon_p-\mu)}\ll 1,
\eeq
where we have used the fact that ${\rm e}^{-\beta\mu}$ is large in the
classical limit.

\end{petitchar}

At this point it is easy to understand the origin of the breakdown of perturbations theory. 
 The critical region is characterized by the fact that  
all the terms in the integrand of
(\ref{LEcl}) become of the same order of magnitude. This occurs for $k \simle k_c$,
with $k_c$ such that:
\beq\label{ppp}
\frac{k_c^2}{2m}\sim \mu'\sim \frac{a}{m} \frac{T}{\mu'} k_c^3,
\eeq
where $(T/\mu')k_c^3$ is the contribution to the density of the modes
with $k\sim
k_c$. From eq.~(\ref{ppp}) we see that $k_c\sim a/\lambda^2$.  For $k\simeq k_c$
perturbation
theory in
$a$ makes no sense, and in fact all terms in the perturbative
expansion are  infrared
divergent. For
$k_c\ll k\ll
\lambda^{-1}$, perturbation theory is applicable. 

\begin{petitchar}
\renewcommand{\baselinestretch}{.90}\small
\noindent{\bf Remark.} 
In a trap the effect of  critical fluctuations is subleading. To see that let us estimate the size of the critical region in a trap \cite{arnold01}. Recall that at the transition the thermal wavelength is $\lambda\sim a_{ho} N^{-1/6}$, where $a_{ho}=1/\sqrt{m\omega_0}$ is the characteristic size of the harmonic oscillator trap  (see the discussion after eq.~(\ref{napproxtrap})), and the size $R_{cl}$ of the thermal cloud where most of the particles sit is $R_{cl}\sim a_{ho}N^{1/6}\sim \lambda N^{1/3}$. According to eq.~(\ref{ppp}), the  critical region is reached when the chemical potential deviates from its critical value by an amount $\simle k_c^2/{2m}\sim a^2/(m\lambda^4)$. In a trap the effective local potential is of the form $\mu-\half m\omega_0^2r^2$. Taking $\mu$ at its critical value, one finds that the particles will be in the critical region as long as $r\simle R_{cr}$ with 
$m\omega_0^2 R_{cr}^2\sim a^2/(m\lambda^4)$, or $R_{cr}\sim a(a_{ho}/\lambda)^2\sim a N^{1/3}$. Thus the relative size of the critical region is $R_{cr}/R_{cl}\sim a/\lambda$, and the ratio of the particles in the critical region to the total number of particles is $\sim (a/\lambda)^3$. Under such conditions, it can be shown that one can use perturbation theory to estimate the corrections due to the interactions to the relation (\ref{ndemuetT}) in a trap; the resulting  contributions to the shift of $T_c$ are then subleading in $a$, as compared to the mean field effect discussed in the first lecture \cite{arnold01}.
\end{petitchar}

In view of the forthcoming discussions, it is convenient to rescale the field $\psi_0$ and to parametrize it in terms of two real fields
$\varphi_1,\varphi_2$:  $\psi_0=\sqrt{mT}(\varphi_1+i\varphi_2)$.  The partition function
then reads
\beq
 {\cal Z}= \int {\cal D}\varphi \,{\rm e}^{-{\cal S}}
 , \label{eONpart}
\eeq
where ${\cal S}(\varphi)= (H-\mu N)/T$ is given by:
\label{eLGWphi}:
\beq
{\cal S} \left( \varphi \right)= \int \left\lbrace{ 1 \over 2} \left[
\partial_{\mu} \varphi (x) \right]^2+{1 \over 2}r
\phib^2 (x)+{u \over 4!} \left[ \varphi^2(x) \right]^2 \right\rbrace \d^{d}x\,,
\label{eactON}
\eeq
where $\varphi^2=\varphi_1^2+\varphi_2^2$ and:
\beq\label{def:rmu}
r=-2mT\mu, \qquad u=96 \pi^2\frac{a}{\lambda^2}.
\eeq
In eq.~(\ref{eactON}) we have kept the dimension $d$ of the spatial
integration arbitrary for the convenience of forthcoming discussions.
The single particle Green's function $G(p)$ is related to the inverse
two-point function $\Gamma^{(2)}(p)$ of the classical field theory by
\beq\label{Ugamma2}
2mT G^{-1}(p)= \Gamma^{(2)}(p),\qquad p^2+U(p)= \Gamma^{(2)}(p)-\Gamma^{(2)}(0).
\eeq

 As it stands this
field theory suffers from UV divergences.  These are absent in the
original theory, the higher frequency modes providing a large momentum
cutoff $\sim  1/\lambda$.  This cutoff may be restored when
needed, but, as we show later, since the shift of the critical temperature is
dominated by long distance properties it is independent of the precise
cutoff procedure.

We shall also find useful to consider the $O(N)$ symmetric generalization of the Euclidean action (\ref{eactON}).  The field $\varphi(x)$
then has $N$ real components, and, e.g.,
\beq
\varphi^2=\sum_{i=1}^N \varphi_i^2\,,
\eeq
and the shift in the critical density is given by
\beq
\Delta n_c&=&2mT \sum_{i=1}^N \left[  \left<\phi_i^2 \right>_{a\ne 0}- \left<\phi_i^2 \right>_{a=0}\right] \nonumber\\&=& {2mT \,
N\,}\int\frac{\d^d p}{(2\pi)^d}\,\left( \frac{1}{\Gamma^{(2)}(p)}-\frac{1}{p^2}\right)\label{eONrho},
\eeq
with $\delta_{ij}/{\Gamma^{(2)}(k)}$  the connected two-point correlation
function.

 The advantage of this generalization is that it provides
us with a tool, the large $N$ expansion, which allows us to calculate at
the critical point (for a recent review see e.g. \cite{Moshe:2003xn}).

\subsubsection{Linear dependence of the density correction}

\label{class}

It is now easy to see the origin of the linear relation between $\Delta n_c$ and $a$. Note first that the action (\ref{eactON}) contains a single dimensionfull parameter, $u$, $r$ being adjusted for any given $u$ to be at criticality. In fact the effective three dimensional theory is ultraviolet divergent, so there is a priori another parameter, the ultraviolet cut-off $\Lambda\sim 1/\lambda$. It follows then from dimensional analysis  that $U(p)$ defined in eq.~(\ref{defU}) can be written as
\beq\label{def:sigma}
U(p=xu)=u^2 \sigma(x,u/\Lambda).
\eeq
Now, the diagrams involved in the calculation of $U$ are ultraviolet convergent, so that $U$ is in fact independent of the cut-off $\Lambda$, and the infinite cut-off limit can be taken. Note however that the validity of the classical field approximation requires that all momenta
involved in the various integrations  are small in comparison
with  $\Lambda \sim \lambda^{-1}$  or, in other words, that the integrands
are  negligibly small for momenta
$k\sim \lambda^{-1}$. Only then  can we ignore the effects of non vanishing Matsubara frequencies and use for instance the approximate
form of the statistical
factors (\ref{BESOFT}). In other words, the infinite cut-off limit is meaningful only if letting the cut-off becoming bigger than $\lambda^{-1}$ does not affect the results. This implies that 
$u\lambda\sim a/\lambda$ is  sufficiently small.

 In the region of validity of the classical field approximation, that is, for small enough $u$, $\sigma(x,u/\Lambda)$ becomes a universal function $\sigma(x)$, independent of $u$, and   $\Delta n_c$ in eq.~(\ref{deltanc2}) takes the form
\begin{equation}
\Delta n_{c}=-\frac{2u}{\pi \lambda ^{2}}\int dx
\frac{\sigma (x)}{x^{2}+\sigma (x)},  \label{cl-6}
\end{equation}
showing that the change in the critical density is indeed linear in $a$.

\subsubsection{Renormalization group argument}

    The linearity of the relation between the shift in $T_c$ and the scattering length can also be understood from a simple renormalization group analysis. Let us introduce a large momentum cutoff $\Lambda
\sim 1/\lambda$, and a dimensionless coupling constant $g$
\beq\label{coupling}
g=\Lambda^{d-4}u\propto \left(a\over \lambda\right)^{d-2}\,.
\eeq
At $T_c$ the two-point function in momentum space satisfies the
renormalization group equation \cite{Zinn-Justin:2002ru}
\beq\label{RGeq}
\left(\Lambda{\partial\over\partial \Lambda}+\beta(g){\partial \over\partial
g}-\eta(g)\right)\Gamma^{(2)}(p,\Lambda,g)=0\,.
\eeq
This equation, together  with dimensional analysis, implies that the two-point
function has the general form
\beq
\Gamma^{(2)}(p,\Lambda,g)=p^2 Z(g)F\bigl(p/k_c\bigr),
\label{f}
\eeq
where $k_c=k_c(g)$ is a function of $g$ which,  on dimensional grounds, is proportional to $\Lambda$, so that $\Lambda\partial_\Lambda k_c=k_c$. The ansatz (\ref{f}) for $\Gamma^{(2)}(p,\Lambda,g)$ provides then a solution of eq.~(\ref{RGeq}) if $Z(g)$) and $k_c(g)$ obey the equations
\beq
{\partial \ln Z(g)\over \partial g}&=&\frac{ \eta(g)}{\beta(g)},
\label{beta}
\\
{\partial \ln k_c (g)\over \partial g}&=& -\frac{1}{\beta(g)} \,.
\label{lam}
\eeq
 \begin{figure}
\includegraphics[scale=.25,angle=0]{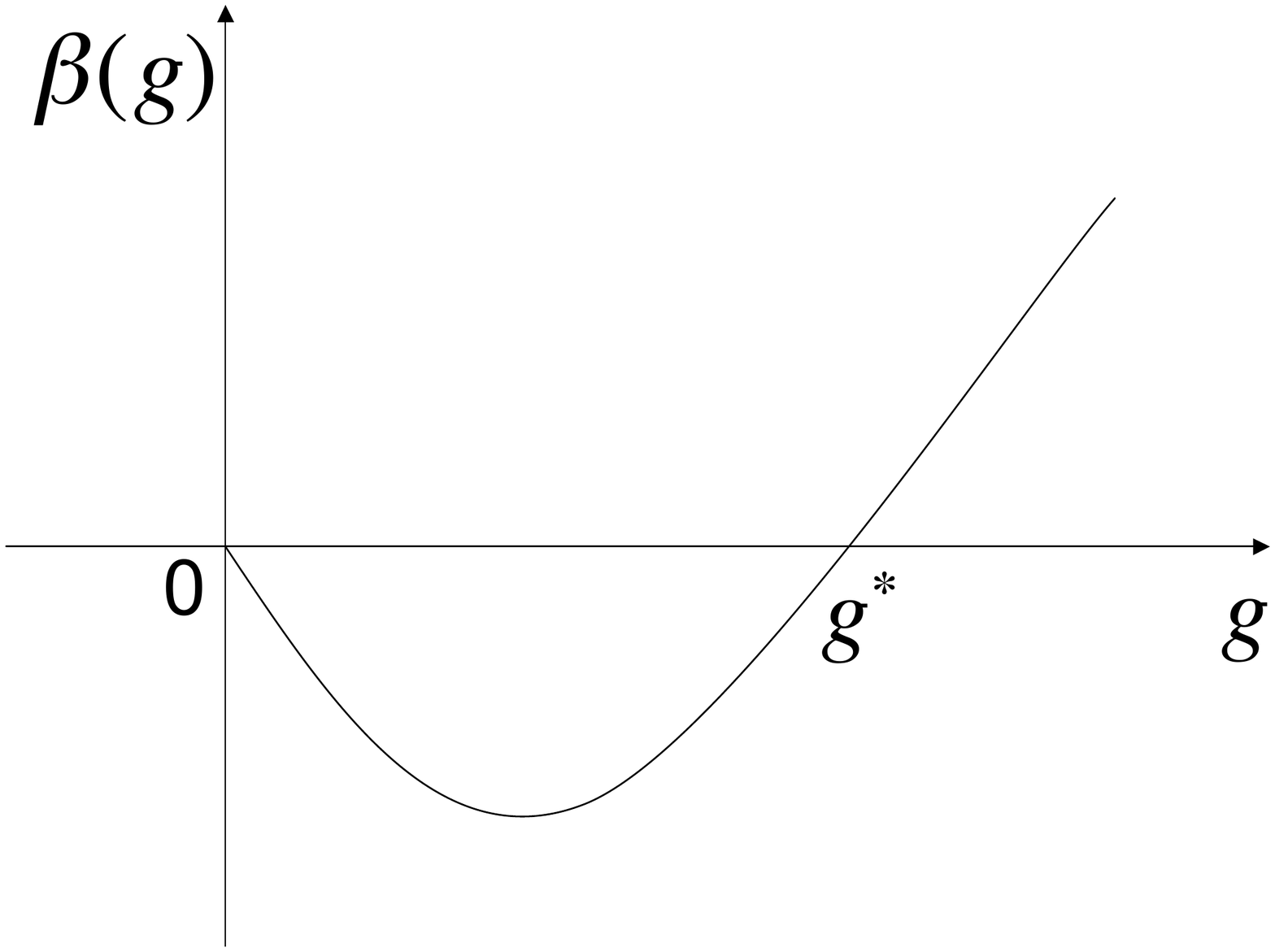}
\begin{center}
\caption{The $\beta$-function with the two fixed points at $g=0$ and $g=g^*$. \label{betafunction}}
\end{center}
\end{figure}
Since $\beta(g)=-(4-d) g +(N+8)g^2/48\pi^2+{\cal O}(g^3)$, $\beta(g)$ is
of order $g$ for small $g$ in $d<4$; similarly
$\eta(g)=(N+2)g^2/(72(8\pi^2)^2)+{\cal O}(g^3)$.  Therefore
\beq
 Z(g)=\exp\int_0^g{\eta(g') \over \beta(g')}\d g'=1+{\cal O}(g^2);
\eeq
to leading order $Z(g)$ =1.  The function $k_c(g)$ is then obtained by
integrating eq.~(\ref{lam}),
\beq
k_c(g)=g^{1/(4-d)}\Lambda\exp\left[-\int_0^g\d g'\left({1\over
\beta(g')} +{1\over (4-d)g'}\right)\right].
\eeq
In $d=3$, 
\beq
k_c(g)=\Lambda \frac{gg^*}{g^*-g},
\eeq
where $g^*$ is the infrared fixed point (see Fig.~\ref{betafunction}).
The scale $k_c(g)$ plays a specific role in the analysis as the
crossover separating a universal long-distance regime, where
\beq\label{IRscaling}
\Gamma^{(2)}(p)\propto p^{2-\eta} \qquad p\ll k_c(g) ,
\eeq
governed by the non-trivial zero, $g^*$, of the $\beta$-function, from a
universal short distance regime governed by the Gaussian fixed point, $g=0$,
where
\beq
\Gamma^{(2)}(p)\propto p^2 \qquad k_c(g) \ll p\ll \Lambda\,.
\eeq
However such a regime exists only if $k_c(g)\ll \Lambda$, i.e., if
there is an intermediate scale between the infrared and the microscopic scales;
otherwise only the infrared behavior can be observed.  In a generic situation $g$ is
of order unity, and thus $k_c(g)$ is of order $\Lambda$, and the universal
large momentum region is absent.  Instead $k_c(g)\ll \Lambda$ implies
that $g$ be small. Since
 $g\sim a/\lambda\ll 1$, see eq.~(\ref{coupling}), this
condition is satisfied in the present situation.

   It is then easy to show, repeating essentially the same analysis as before,  that with this condition, $\Delta T_c\propto k_c(g)$. We set $p=x k_c(g)$, and find for the integral in eq.~(\ref{eONrho}) the general
form
\beq\label{deltarho}
\int\frac{\d^d p}{(2\pi)^d}\,\left( \frac{1}{\Gamma^{(2)}(k)}-\frac{1}{p^2}\right)=k_c(g)\int \frac{\d^3 x}{(2\pi)^3} {1\over
  x^2}\left( {1\over F(x)} -{ 1}\right);
\eeq
the $g$ dependence is entirely contained in $k_c(g)$, and for small $g$, $k_c(g)\simeq u$.

Recall  that both the
perturbative large momentum region and the non-perturbative infrared region
contribute to the integrand in eq.~(\ref{deltarho}), or equivalently  in eq.~(\ref{cl-6}), and that the functions $F(x)$ or $\sigma(x)$ cannot be calculated using perturbation theory.
  
The $1/N$ expansion allows an explicit calculation, and yields, in leading order  for the coefficient $c$ in eq.~(\ref{deltaTc}) the value  $c=2.3$ \cite{BigN}. However, the best numerical estimates  for $c$  are
those which have been obtained using the lattice technique by two groups, with the
results: $c=1.32\pm0.02$
\cite{latt2} and $c=1.29\pm 0.05$ \cite{latt1}. The availabilty
of these results has turned the
calculation of $c$ into a testing ground for other non perturbative methods:
expansion in  $1/N$
\cite{BigN,Arnold:2000ef}, optimized perturbation theory \cite{Kneur04,souza},
  resummed    perturbative  calculations to high loop orders
\cite{Kastening:2003iu}. Note that while the
latter methods yield critical exponents with several significant
digits, they predict $c$ with only a 10\%
accuracy. This illustrates the difficulty of getting an accurate
determination of $c$ using  (semi) analytical
techniques.

\begin{petitchar}
\renewcommand{\baselinestretch}{.90}\small
\noindent{\bf Remark.} 
The linear dependence in $a$ of the shift of $T_c$ holds only if $a$ is small enough, as we have already indicated. When $a$ is not small enough various corrections need to be taken into account that alter the simple linear law. In particular, corrections come from the non vanishing Matsubara frequencies, and their impact on the effective theory for $\psi_0$. Such corrections have been analyzed in detail in \cite{Arnold:2001nn}
 (see also \cite{NEW}). The net result is the following expression for $\Delta T_c$:
\beq
\frac{\Delta T_c}{T_c^0}= c (an^{1/3})+\left[ c_2'\ln(an^{1/3})+c_2''\right] (an^{1/3})^2+{\cal O}(a^3n).
\eeq
Aside from such corrections, the evaluation of $\Delta T_c$ in higher order requires that one improves the  accuracy of the effective hamiltonian and include for instance effective range corrections. 
\end{petitchar}

%%%%%%%%%%%%%%%%%%%%%%%%%%%%%%%%%%%%%%%%%%%%%%%%%%
%%%%%%%%%%%%%%%%%%%%%%%%%%%%%%%%%%%%%%%%%%%%%%%%%%

\section{LECTURE 3 : The Non Perturbative Renormalization Group and the calculation of $c$}

The analysis of the previous lectures has shown that the coefficient  $c$ in the formula (\ref{deltaTc}) can be written as:
 \beq\label{formulac2}
 c\,\equiv -\frac{256
\pi^3}{\left(\zeta(3/2)\right)^{4/3}} \,\frac{\Delta\langle\varphi_i^2\rangle}{Nu},\eeq 
where $\Delta\langle\varphi_i^2\rangle$ represents the change, due to interactions, of the fluctuations of
the  scalar field of the $O(N)$ symmetric model, and is given by:
\beq\label{integralc}
\frac{\Delta\langle \varphi_i^2\rangle}{Nu}=- \int\frac{{\rm }dx}{2\pi^2}\,
\frac{\sigma(x)}{x^2+\sigma(x)},
\eeq
where $\sigma(x)=u^{-2}U(p=xu)$ and $U(p)$ is essentially  the self-energy of the field at  criticality (see eqs.~(\ref{eONrho}) and (\ref{Ugamma2})). In order to get the numerical factor in eq.~(\ref{formulac2}), we have combined eqs.~(\ref{deltaTc}), (\ref{deltaTdeltan}), (\ref{criticalline}) and (\ref{def:rmu}). Eq.~(\ref{formulac2}) has been written for arbitray $N$ in order to be able to compare with  all available results. Bose-Einstein condensation corresponds to $N=2$.

\begin{figure}
\includegraphics[scale=.45,angle=0]{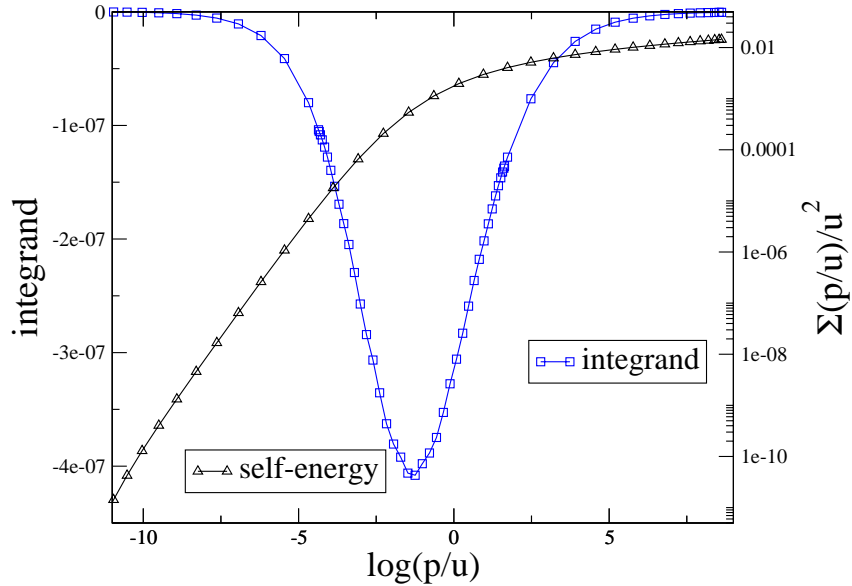}
\begin{center}
%\onefigure[scale=.250,angle=-90]{curvasuperpuesta}
\caption{The function $\sigma(x=p/u)$ at criticality,  and the integrand of eq.~( \ref{integralc}) as a function of
$\ln(p/u)$, for $u\simle 10^{-4}$ for which $\sigma(x)$ is independent of $u$\label{fig:sigma}. From Ref.~\cite{Blaizot:2004qa}.}
\end{center}
\end{figure}

The integrand of eq.~(\ref{integralc}),  calculated in the approximate scheme described in \cite{Blaizot:2004qa}, is shown in Fig.~\ref{fig:sigma}.
The  minimum
occurs at  the typical scale $k_c$ which separates the scaling region  from  the high
momentum region where perturbation theory applies. As we have already emphasized, the difficulty in getting a precise evaluation of the integral
(\ref{integralc}) is
that it requires an  accurate determination of $\sigma(x)$ in a large
region of momenta including in particular the crossover region between two
different physical regimes; as we have seen in the previous lecture, this cannot be done using perturbation theory, eventhough the coupling constant, $\sim a$, is small. 

In this last lecture, I shall show how  the non perturbative renormalization group
(NPRG) can be used to calclate $\sigma(x)$. The NPRG \cite{Wetterich93,Ellwanger93,Tetradis94,Morris94,Morris94c}  (sometimes called ``exact'' or ``functional'', depending on which aspect of the formalism one wishes to emphasize)  stands out as a
 promising tool, suggesting
 new approximation schemes which are not easily formulated in other,
more conventional, approaches in field theory or many body
physics. It has been applied successfully to a variety of
 problems, in condensed matter, particle or nuclear
physics (for  reviews, see e.g.
\cite{Bagnuls:2000ae,Berges02,Canet04}). In most of these problems
however, the focus is on long wavelength modes  and  the solution
of the NPRG equations involves generally a derivative expansion
which only allows for the determination of the $n$-point functions
and their derivatives essentially only at vanishing external momenta. This is
not enough in the present situation where, as we have seen,  a full knowledge of the momentum dependence of the
2-point  function is required. We have therefore developed new methods to solve the NPRG equations. The following sections will briefly present some of the ingredients involved, without going into the technical details of their implementation. These can be found in Refs.~\cite{Blaizot:2004qa,Blaizot:2005wd,Blaizot:2006vr,Blaizot:2005xy,Blaizot:2006ak}, from which much of  the material presented here is borrowed. For definiteness, we  focus on the $O(N)$ symmetric scalar field theory with action (\ref{eactON}). See also the lecture by H. Gies in this volume \cite{Gies:2006wv} for further introduction to these techniques, and their application to other theories, in particular non-abelian gauge theories. Other applications of the renormalization group to Bose-Einstein condensation can be found in \cite{Stoof2,Andersen:1998bc,Andersen:1999dy}.

 \subsection{The NPRG equations}

The NPRG   allows the construction of a set of
effective actions $\Gamma_\kappa[\phi]$ which interpolate between the classical action $S$ and the full effective action $\Gamma[\phi]$: In
$\Gamma_\kappa[\phi]$   the magnitude of long wavelength
fluctuations of the field is controlled by an infrared regulator
depending on a continuous parameter $\kappa$ which has the dimension of a momentum.  The full effective action is obtained for the value $\kappa=0$, the situation with no infrared cut-off. In the other limit, corresponding to a value of $\kappa$ of the order of the  microscopic scale $\Lambda$ at  which fluctuations are suppressed, $\Gamma_{\kappa=\Lambda}[\phi]$ reduces to the classical action \footnote{Note that depending on the choice of the regulator, not all fluctuations may be suppressed when $\kappa=\Lambda$. However, for renormalisable theories, and if $\Lambda$ is large enough, the effects of these remnant fluctuations can be absorbed into a redefinition of the parameters of the classical action.}.

In practice the fluctuations are controlled by adding to the classical  action (\ref{eactON})
the $O(N)$ symmetric regulator
 \beq\label{regulaction}
  \Delta S_\kappa[\varphi] =\frac{1}{2} \int \frac{{\rm d}^dq}{(2\pi)^d}
\, \varphi_i(q)\,R_\kappa(q) \,\varphi_i(-q),
  \eeq\normalsize
where $R_\kappa$ denotes a family of ``cut-off functions''
depending  on  $\kappa$. As we just said, the role of $\Delta S_\kappa$ is to
suppress the fluctuations with momenta $q\simle \kappa$, while
leaving unaffected those with $q\simge \kappa$. Thus,
typically, $ R_{\kappa}(q)\to\kappa^2$ when $ q \ll \kappa$, and
$R_{\kappa}(q)\to 0$
 when $ q\simge \kappa$.
There is a large freedom in the choice of $R_\kappa(q)$,
abundantly discussed in the literature
\cite{Ball95,Comellas98,Litim,Canet02}. The choice of the cut-off function matters when approximations are done, as is the case in all situations of practical interest. Most of the results that will be presented here have been obtained with the cut-off
function proposed in \cite{Litim}:
\beq\label{reg-litim}
R_\kappa(q)\propto (\kappa^2-q^2) \theta(\kappa^2-q^2).
\eeq
This regulator has the advantage of allowing  some calculations to be done analytically. 

\begin{figure}[t!]
\begin{center}
\includegraphics*[scale=0.6,angle=0]{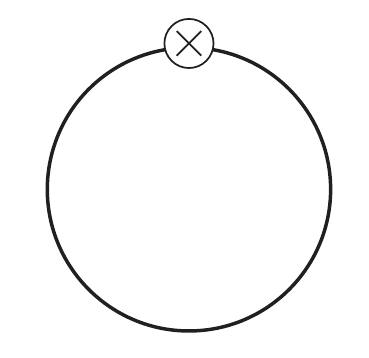}
\end{center}
\caption{ Diagrammatic illustration of the r.h.s. of the  flow equation
of the effective action, eq.~(\ref{mastereq}).
The crossed circle represents
an insertion of $\del_\kappa R_\kappa$, and the thick line a full
propagator in an arbitrary background field.\label{fig:mastereq}}
\end{figure}

For each value of $\kappa$, one defines the generating functional of
connected Green's functions \beq W_\kappa[J]= \ln \int D\varphi
\hspace{.1cm} \exp\left\{  {-S[\varphi]-\Delta S_\kappa[\varphi]+\int
d^dx \varphi(x)J(x)}\right\} . \eeq
The Feynman diagrams contributing to $W_\kappa$ are those of
ordinary perturbation theory, except that the propagators contain
the infrared regulator.  We also define the effective action,
through a modified Legendre transform that includes the  explicit subtraction of $\Delta S_\kappa$:  \beq
\Gamma_\kappa[\phi]=-W_\kappa[J_{\phi}]+\int d^dx \hspace{.1cm}
\phi(x) J_{\phi}(x)-\Delta S_\kappa[\phi], \eeq where $J_\phi$ is
obtained by inverting the relation
\beq
\label{Legendre1} \phi_{\kappa,J}(x)\equiv\left\langle \varphi(x)
\right\rangle_{\kappa,J} =\frac{\delta W_\kappa}{\delta J(x)}.
\eeq
for $\phi_{\kappa,J}(x)=\phi$. Note that, in this
inversion, $\phi$ is considered as a given variable, so that
$J_\phi$ becomes  implicitly dependent on $\kappa$. Taking this in to account, it is easy to derive the following exact flow equation for  $\Gamma_\kappa[\phi]$:
\beq\label{mastereq}
\partial_\kappa \Gamma_\kappa[\phi]=\frac{1}{2}{\rm tr}\int \frac{d^dq}{(2\pi)^d}
\,\partial_\kappa R_\kappa(q)\, G(\kappa,q), 
\eeq
where the trace ${\rm tr}$ runs over the O($N$) indices, and 
\beq\label{Ggamma2}
G^{-1}_{ij}(q;\kappa;\phi)=\Gamma^{(2)}_{ij}(\kappa;q;\phi)+\delta_{ij} R_\kappa(q).
\eeq
with $\Gamma^{(2)}$  the second functional derivative of $\Gamma_\kappa$
with respect to $\phi$ (see eq.~(\ref{rednptfcts}) below). Eq.~(\ref{mastereq}) is
the master equation of the NPRG. Its solution  yields the effective action $\Gamma[\phi]=\Gamma_{\kappa=0}[\phi]$ starting with the initial condition $\Gamma_{\kappa=\Lambda}[\phi]=S[\phi]$. Its right hand side has the
structure of a one loop integral, with one insertion of
$\partial_\kappa R_\kappa(q^2)$ (see Fig.~\ref{fig:mastereq}). This simple structure should not hide the fact that   eq.~(\ref{mastereq}) is an exact equation ($G$ in the r.h.s; is the exact propagator), and as such it is of limited use unless some approximation is made. Before we turn to such approximation, let us further analyze the content of eq.~(\ref{mastereq}) in terms of the $n$-point functions. 

\begin{figure}
\begin{center}
\includegraphics[scale=.3] {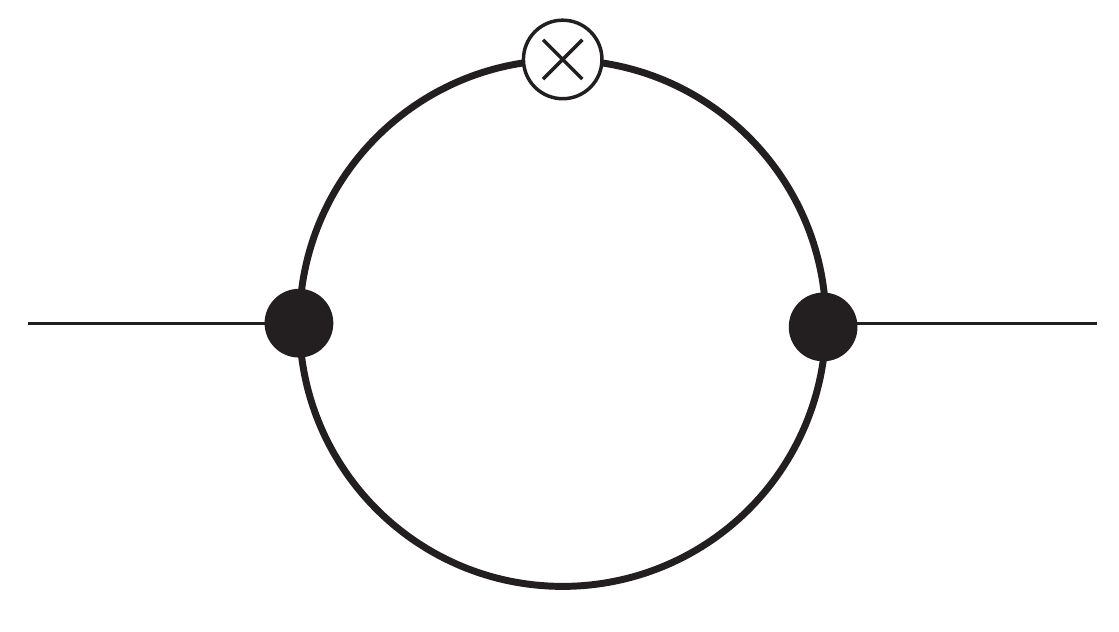} \hspace{20mm}
\includegraphics[scale=.3] {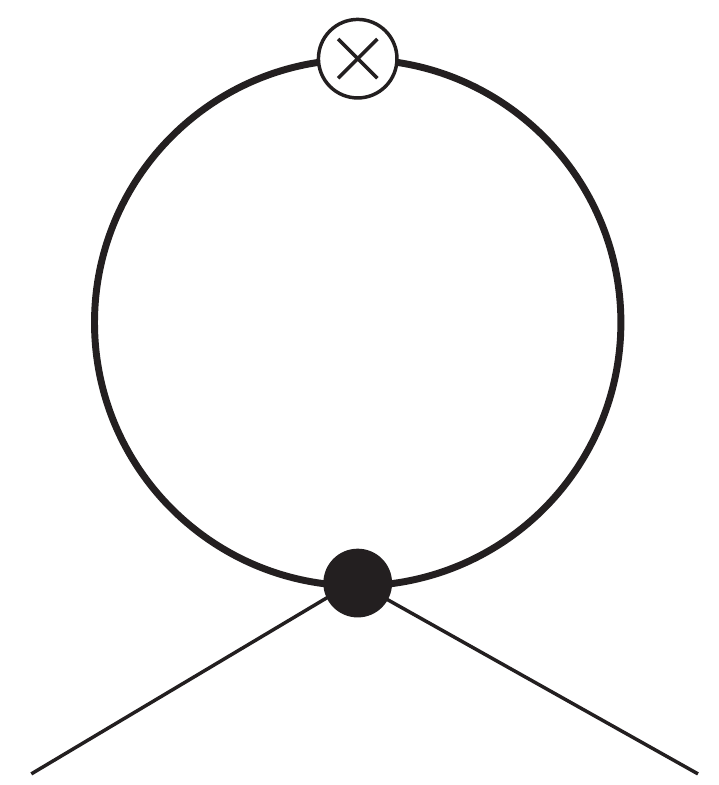}
\end{center}
\caption{The two diagrams contributing to the flow of the 2-point function in a constant background field. The lines represent dressed propagators, $G_\kappa$. The
 cross represents an insertion of $\partial_\kappa R_k$. The vertices denoted by   black dots are $\Gamma^{(3)}_\kappa$ and $\Gamma^{(4)}_\kappa$.  When the background field vanishes, so does the diagram on the left. \label{2-point-diagrams}}
\end{figure}

As well known (see e.g. \cite{Zinn-Justin:2002ru}), the effective action
$\Gamma[\phi]$ is the generating functional of the one-particle
irreducible $n$-point functions. This property extends trivially
to $\Gamma_\kappa[\phi]$:
\beq\label{rednptfcts}
\Gamma_\kappa[\phi]= \sum_n\frac{1}{n!}  
\int d^dx_1\dots\int d^dx_{n} \,\Gamma_\kappa^{(n)}[\phi;x_1,\cdots, x_n] \,\phi(x_1) \dots
 \phi(x_n). \nonumber\\ \eeq 
 By differentiating
eq.~(\ref{mastereq}) with respect to $\phi$,   letting the
field be constant, and taking a Fourier transform, one gets the flow equations for all $n$-point
functions in a  constant background field $\phi$. For instance, the
flow of the 2-point
function in a constant external field reads:
\begin{eqnarray}
\label{gamma2champnonnul}
\partial_\kappa\Gamma_{ab}^{(2)}(p,-p;\kappa;\phi)&=&\int
\frac{d^dq}{(2\pi)^d}\partial_\kappa R_\kappa(q)\left\{G_{ij}(q;\kappa;\phi)\Gamma_{ajk}^{(3)}(p,q,-p-q;\kappa;\phi)\right. \nonumber \\
&&\times G_{kl}(q+p;\kappa;\phi)\Gamma_{blm}^{(3)}(-p,p+q,-q;\kappa;\phi)G_{mi}(q;\kappa;\phi) \nonumber \\
&&\left.-\frac{1}{2}G_{ij}(q;\kappa;\phi)\Gamma_{abjk}^{(4)}(p,-p,q,-q;\kappa;\phi)G_{ki}(q;\kappa;\phi)\right\}.\nonumber\\ 
\end{eqnarray}
The corresponding diagrams contributing to the flow are shown
in Fig.~\ref{2-point-diagrams}.

\begin{figure}[t!]
\begin{center}
\includegraphics*[scale=0.6,angle=0]{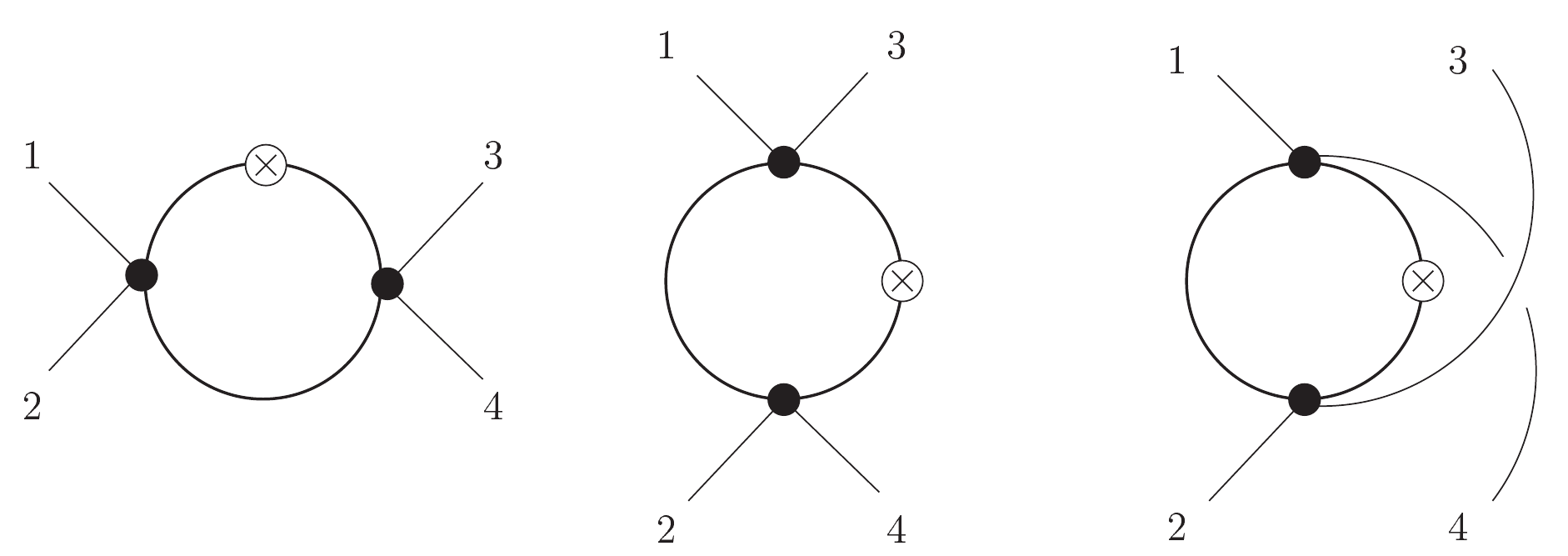}
\end{center}
\caption{\label{fig:dGamma4} Diagrammatic illustration of the r.h.s. of the flow
equation for the 4-point function, eq.~(\ref{4point}): contribution of the 4-point functions (represented by black disks) in the three possible channels, from left to right. The crossed circle represents
an insertion of $\del_\kappa R_\kappa$, and the thick line a full
propagator.\label{fig:4point}}
\end{figure}

\begin{figure}[t!]
\begin{center}
\includegraphics*[scale=0.6,angle=0]{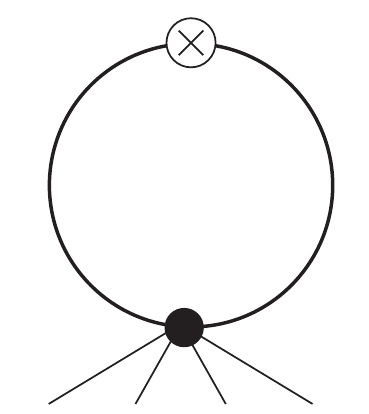}
\end{center}
\caption{\label{fig:dGamma4_6} Diagrammatic illustration of the r.h.s. of the flow
equation for the 4-point function, eq.~(\ref{4point}):  contribution of the 6-point function $\Gamma^{(6)}$ (represented by a black disk). The crossed circle represents
an insertion of $\del_\kappa R_\kappa$, and the thick line a full
propagator.\label{fig:6point}}
\end{figure}

 When the external field vanishes, this equation simplifies greatly since then $\Gamma^{(3)}$ vanishes, and $\Gamma^{(2)}$ is diagonal :
 \beq\label{SigmaON}
 \Gamma^{(2)}_{ab}(\kappa;q)=\delta_{ab}(q^2+\Sigma(\kappa;q),
 \eeq
 which defines the self-energy $\Sigma$ (this $\Sigma$ differs from $\Sigma_{cl}$ introduced in sect.~3.2 by a factor $2m$; see e.g. eqs.~(\ref{defU}) and (\ref{Ugamma2})). We get then
\beq\label{eq:dGamma2}
 \hspace{-1cm}\partial_\kappa
\Gamma^{(2)}_{ab}(\kappa;p)=-\frac{1}{2}\int
\frac{d^dq}{(2\pi)^d} \,\partial_\kappa R_\kappa(q)\,G^2(\kappa;q)
\,\Gamma^{(4)}_{abll}(\kappa;p,-p,q,-q),
\eeq
where 
\beq \label{SigmaON2}G^{-1}(\kappa,q)=q^2+R_\kappa(q)+\Sigma(\kappa;q). \eeq
Similarly, the flow of the  the 4-point function in vanishing field reads:
\begin{eqnarray}
\label{4point}
&&\partial_\kappa\Gamma^{(4)}_{abcd}(\kappa;p_1,p_2,p_3,p_4)=
\int \frac{d^dq}{(2\pi)^d}\,\partial_\kappa R_k(q^2)\,G^2(\kappa;q) \nonumber \\
&&\hspace{.5cm}\times\left\lbrace
G(\kappa;q+p_1+p_2)\Gamma^{(4)}_{abij}(\kappa;p_1,p_2,q,\cdot)
\Gamma^{(4)}_{cdij}(\kappa;p_3,p_4,-q,\cdot) \right.\nonumber \\
&&\hspace{.80cm}+\,G(\kappa;q+p_1+p_3)\Gamma^{(4)}_{acij}(\kappa;p_1,p_3,q,\cdot)
\Gamma^{(4)}_{bdij}(\kappa;p_2,p_4,-q,\cdot) \nonumber \\
&&\hspace{.80cm}\left.+\,G(\kappa;q+p_1+p_4)\Gamma^{(4)}_{adij}(\kappa;p_1,p_4,q,\cdot)
\Gamma^{(4)}_{cbij}(\kappa;p_3,p_2,-q,\cdot) \right\rbrace \nonumber \\
&&\hspace{2cm}-\,\frac{1}{2}\int
\frac{d^dq}{(2\pi)^d}\partial_\kappa R_\kappa(q)G^2(\kappa;q)
\Gamma^{(6)}_{abcdii}(\kappa;p_1,p_2,p_3,p_4,q,-q) , \nonumber\\ 
\end{eqnarray}
 where we have used the abbreviated notation $\Gamma^{(4)}_{abij}(\kappa;p_1,p_2,q,-p_1-p_2-q)$ $\to$ $\Gamma^{(4)}_{abij}(\kappa;p_1,p_2,q,\cdot)$.
The four contributions in the r.h.s.~of eq.~(\ref{4point}) are represented in the
diagrams shown in figs.~\ref{fig:dGamma4} and \ref{fig:dGamma4_6}.

Eqs.~(\ref{eq:dGamma2}) and (\ref{4point}) for the 2- and 4-point
functions constitute the beginning of an infinite hierarchy of exact
equations for the $n$-point functions, reminiscent of the Schwinger-Dyson hierarchy, with the flow equation for
the $n$-point function involving all the $m$-point functions up to
$m=n+2$.  Clearly, solving this hierarchy requires approximations. A most natural approximation would rely on a truncation, in order to close the infinite hierarchy, as commonly done for instance in solving  the Schwinger Dyson equations. For instance, one could ignore in eq.~(\ref{4point})  the effect of the 6-point function on the flow of the 4-point function, arguing for instance that the 6-point function is perturbatively of order $u^3$. However such truncations prove to be not accurate enough for the present problem.    Besides, the NPRG offers the possibility of exploring new approximation schemes that involve the entire functional $\Gamma[\phi]$ rather than the individual $n$-point functions. An example of such an approximation is provided by the derivative expansion described in the next section.

\subsection{The local potential approximation}

\label{LPA}

The derivative expansion  
exploits the fact that the regulator in the flow
equations (e.g. eqs.~(\ref{eq:dGamma2}) or (\ref{4point})) forces
the loop momentum $q$ to be smaller than $\kappa$, i.e., only
momenta $q\simle \kappa$ contribute to the flow. Besides, in general, the
regulator insures  that   all vertices are smooth functions of
momenta. Then, in the calculation of the $n$-point functions at
vanishing external momenta $p_i$, it is  possible to expand the $n$-point functions
in the r.h.s. of the flow equations in terms of $q^2/\kappa^2$, or equivalently in terms of the derivatives of
the field.

 In leading order, this procedure reduces to the so-called local potential approximation (LPA), which assumes that the effective action has the form:
\begin{equation}\label{gammaLPA}
\Gamma_\kappa^{LPA}[\phi]=\int d^dx \left\{
\frac{1}{2}\partial_{\mu}\phi_i\partial_{\mu}\phi_i+V_\kappa(\rho)\right\},
\end{equation}
where $\rho\equiv \phi_i\phi_i/2$.
The derivative term here is simply the one appearing in the
classical action, and $V_\kappa(\rho)$ is the effective potential.
The exact flow equation for $V_\kappa$ is
easily obtained  by assuming that the field $\phi$ is constant in
eq.~(\ref{mastereq}). It reads: 
 \beq\label{eq-pot}
\partial_\kappa V_\kappa(\rho)=\frac{1}{2}\int
\frac{d^dq}{(2\pi)^d} \partial_\kappa R_\kappa(q) \left\lbrace
(N-1)G_T(\kappa;q)+G_L(\kappa;q)\right\rbrace ,
\eeq
where $G_T$ and $G_L$ are, respectively, the transverse and 
longitudinal   components of the propagator:
\beq\label{translong}
G_{ij}(\kappa;q)=G_T(\kappa;q)\left(\delta_{ij}-\frac{\phi_i\phi_j}{2\rho}\right)
+G_L(\kappa;q)\frac{\phi_i\phi_j}{2\rho}.
\eeq
By using the LPA effective action, eq.~ (\ref{gammaLPA}), one gets
\begin{eqnarray}\label{translong2}
G_T^{-1}(\kappa;q)&=& q^2+V'(\rho)+R_k(q)  ,\nonumber \\
G_L^{-1}(\kappa;q)&=& q^2+V'(\rho)+2\rho V''(\rho)+R_\kappa(q),
\end{eqnarray}
with $V'(\rho)=dV/d\rho$ and $V''(\rho)=d^2V/d\rho^2$.

The  solution of the LPA is well
documented in the literature (see e.g. \cite{Berges02,Canet02}).
In this section, we shall just,  for illustrative purposes, solve approximately these equations, by keeping only a few terms in the expansion of the effective potential in powers of $\rho$ (thereby effectively implementing a truncation which ignores the effect of higher $n$-point functions on the flow). 
The  derivatives of $V_\kappa(\rho)$ with respect to $\rho$ give
the $n$-point functions at  zero external momenta  as a function of
$\kappa$. We shall introduce a special notation for these $n$-point
functions in vanishing external field: \beq\label{defcst} m_\kappa^2
\equiv \left.\frac{dV_\kappa}{d\rho}\right|_{\rho=0} , \hskip 1 cm
g_\kappa\equiv\left.\frac{d^2V_\kappa}{d\rho^2}\right|_{\rho=0} ,
\hskip 1 cm h_\kappa
\equiv \left.\frac{d^3V_\kappa}{d\rho^3}\right|_{\rho=0} . \eeq

The equations for these $n$-point functions are obtained by differentiating once and twice eq.~(\ref{eq-pot}) with
respect to $\rho$, then setting $\rho=0$, and using the definitions
in eq.~(\ref{defcst}). One gets,  respectively: \beq\label{preMk}
\kappa\partial_\kappa m_\kappa^2 = -\frac{(N+2)}{2} g_\kappa
I_d^{(2)} , \eeq and \beq\label{preFk} \kappa\partial_\kappa
g_\kappa = (N+8) g_\kappa^2 I_d^{(3)}(\kappa) - \frac{1}{2} (N+4)\;
h_\kappa I_d^{(2)}(\kappa) , \eeq where we have defined
\beq\label{Ink} I_d^{(n)}(\kappa) &\equiv &\int
\frac{d^dq}{(2\pi)^d}\kappa\partial_\kappa R_\kappa(q^2)G^n(\kappa; q)\nonumber\\
 &= &2 K_d \frac{\kappa^{d+2}}{(\kappa^2+ m^2_\kappa)^n}
 , \eeq the explicit
form in the second line being obtained for the  regulator (\ref{reg-litim}) and
\beq
K_d^{-1}\equiv\;
2^{d-1}\; \pi^{d/2} \; d\;\Gamma(d/2).
\eeq
We have used the fact that  for
vanishing external field, the propagator is diagonal  and $\Sigma(\kappa,q)=m_\kappa^2$ (see Eqs.~(\ref{SigmaON}) and (\ref{SigmaON2})). Eqs.~(\ref{preMk}) and (\ref{preFk}) are solved starting from the initial condition at
$\kappa=\Lambda$:
\beq\label{relclas}
 m_\Lambda^2 =r\qquad   g_\Lambda = \frac{u }{3}.
\eeq

In order to factor out the  large variations which arise when
$\kappa$ varies from the microscopic scale $\Lambda$ to the
physical scale $\kappa=0$, and also to exhibit the fixed point structure,  it is convenient to isolate  the
explicit scale factors and to   define dimensionless quantities:
\beq\label{adimcons}
m_\kappa^2\equiv   \kappa^2\, \hat
m_\kappa^2, \hskip 1 cm g_\kappa \equiv K_d^{-1}  
\kappa^{4-d}\, \hat g_\kappa, \hskip 1 cm h_\kappa \equiv  K_d^{-2}
  \kappa^{6-2d} \,\hat h_\kappa.
\eeq
 If one assumes that $\hat m_\kappa\ll 1$, and ignore the  contribution of $h_\kappa$, then the equation for $\hat g_\kappa$ becomes:
\begin{eqnarray}\label{LPAapprox}
\kappa\frac{d \hat g_\kappa}{d\kappa}=(d-4)\,\hat g_\kappa+  2(N+8) \,\hat g_\kappa^2, \
\end{eqnarray}
and can be solved explicitly:
\beq\label{eq:hatg} \hat g_\kappa=\frac{\hat
g^*}{1+\left(\frac{\kappa}{\kappa_c}\right)^{4-d}} , \eeq where
$\hat g^*$ is the value of $\hat g$ at the IR fixed point,  $\hat
g^*=(4-d)/{(2(N+8))}$,  and $\kappa_c$ the value of $\kappa$ for
which $\hat g_\kappa=\hat g^*/2$. We have: \beq\label{kappac}
\left(\frac{\kappa_c}{\Lambda}\right)^{d-4}=\frac{\hat g^*-\hat
g_\Lambda}{\hat g_\Lambda} \approx\frac{\hat g^*}{\hat g_\Lambda}
. \eeq where the last approximate equality holds if $\hat g^*\gg \hat
g_\Lambda$. In this regime, one recovers the qualitative feature already discussed in the previous lecture: there exists a well defined scale $\kappa_c\ll \Lambda$,  $\kappa_c^{4-d}=uK_d/(3 \hat g^*)$, that  separates the
scaling region, dominated by the IR fixed point, where $\hat g= \hat g^*$, from the
perturbative region,  dominated by the UV fixed point $\hat g=0$
(when $\kappa\gg \kappa_c$, one can expand $\hat g_\kappa$ in
powers of $\kappa_c/\kappa$; in leading order
$g_\kappa=(u/3)(1-(\kappa_c/\kappa)^{4-d})$).

The local potential approximation, or a refined version of it, enters in an essential way in the approximation scheme that we have developed in order to calculate $\Sigma(p)$. Further insight can be gained by studying the $n$-point functions in the large $N$ limit, to which we now turn. 

\subsection{Correlation functions in the  large $N$ limit}

\label{largeN}
 
 The LPA, as well as the  higher orders of the derivative expansion,   give accurate results for the
correlation functions and their derivatives only at zero external
momenta. In order to get insight into the effect of non vanishing
external momenta we consider now  the correlation functions in the
large $N$ limit (at fixed $uN$) \cite{Ellwanger94,Tetradis95,D'Attanasio97,Blaizot:2005xy}. 

For vanishing field, the inverse propagator has the same  form as in the LPA, eq.~(\ref{SigmaON2}), 
where the running mass $m_\kappa$ is here given by a gap equation
\beq
\label{gap}
m_\kappa^2=r+\frac{Nu}{6}\int \frac{d^dq}{(2\pi)^d}\,
(G(\kappa;q)-G(\Lambda;q)).
\eeq
\begin{figure}[t!]
\begin{center}
\includegraphics*[width=10cm]{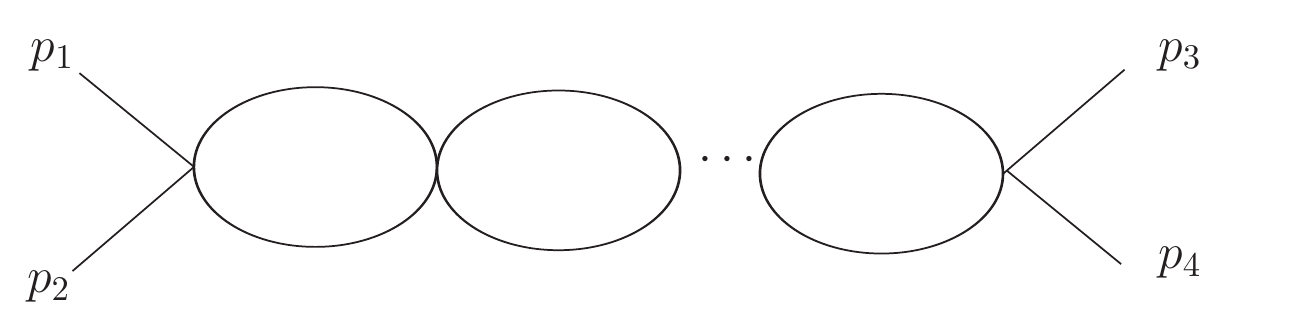}
\end{center}
\caption{\label{coupling4N} Diagrams representing the contribution to  $\Gamma_{1234}^{(4)}(\kappa;p_1,p_2,p_3,p_4)$ in the first channel of eq.~(\ref{4pointlargeN}). This diagram represents also a typical contribution to the function $g_\kappa(p)$ of eq.~(\ref{B3}).  External momenta are counted as incoming momenta, so that $p_1+p_2$ flows into the bubble chain, and $p_3+p_4=-(p_1+p_2)$.}
\end{figure}
The 4-point function has the following structure:
\beq\label{4pointlargeN}
&&\Gamma_{1234}^{(4)}(\kappa;p_1,p_2,p_3,p_4)=\nonumber\\
& &\qquad\delta_{12}\delta_{34}g_\kappa(p_1+p_2)
+\delta_{13}\delta_{24}g_\kappa(p_1+p_3)+\delta_{14}\delta_{23}g_\kappa(p_1+p_4),\nonumber\\
\eeq
where
 $g_\kappa(p)$ is given by
\begin{equation}\label{B3}
g_\kappa(p)=\frac{u}{3} \frac{1}{1+\frac{Nu}{6}B_d(\kappa;p)},
\end{equation}
with
\beq\label{bubble}
B_d(\kappa;p)\equiv \int\frac{d^dq}{(2\pi)^d}\,
G(\kappa;q)G(\kappa;p+q).
\eeq
Finally  the 6-point function $\Gamma_{1234mm}^{(6)}(\kappa; p_1,p_2,p_3,p_4,q,-q)$ (summation over repeated indices is understood) is of the form
\begin{eqnarray}\label{gamma6N}
&&\hspace{-.5cm} \frac{1}{N}\Gamma_{1234mm}^{(6)}(\kappa; p_1,p_2,p_3,p_4,q,-q)\nonumber\\
&&=  h_\kappa(p_1+p_2)\delta_{12}\delta_{34}
+h_\kappa(p_1+p_3)\delta_{13}\delta_{24}+h_\kappa(p_1+p_4)\delta_{14}\delta_{23},
\end{eqnarray}
with
\beq\label{Hkappa}
h_\kappa(p)=N g_\kappa(0)g^2_\kappa(p)\int\frac{d^dq'}{(2\pi)^d}G^2(\kappa;q')G(\kappa;q'+p).
\eeq
\begin{figure}[t!]
\begin{center}
\includegraphics*[width=5cm]{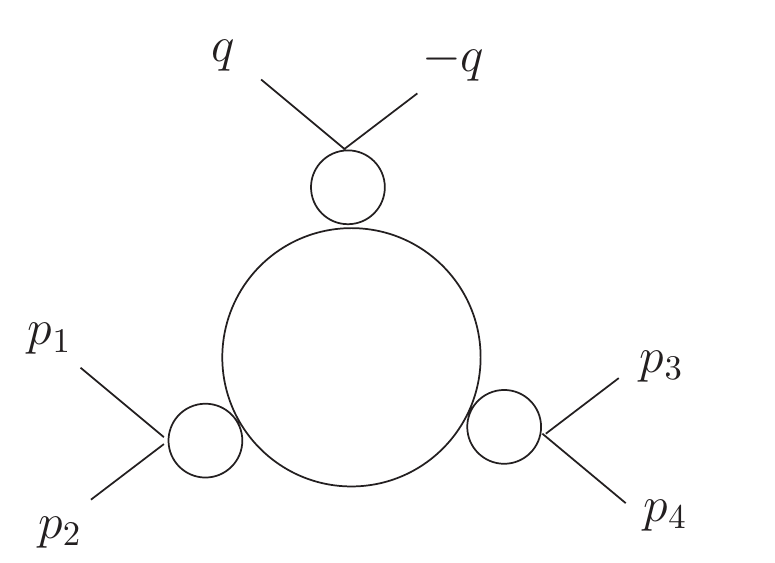}
\end{center}
\caption{\label{coupling6N}Diagrams representing the contribution to $\Gamma_{1234mm}^{(6)}(\kappa; p_1,p_2,p_3,p_4,q,-q)$ in eq.~(\ref{gamma6N}). The small circles represent the function $g_\kappa(p)$ with $p$ the momentum flowing through the vertex. The convention for momenta is as in Fig.~\ref{coupling4N}.}
\end{figure}
Note that a single function, $g_\kappa(p)$, suffices to calculate all the $n$-point functions in the large $N$ limit. These  can be obtained in a straightforward fashion by calculating the corresponding Feynman diagrams with a regulator (see Figs.~\ref{coupling4N} and \ref{coupling6N}). It is however easy to  verify that the various $n$-point functions that we have just written are indeed   solutions of the flow equations in the large $N$ limit.

 To this aim, one notes first that eq.~(\ref{eq:dGamma2}) reduces to an equation for the running mass:
\beq \label{deriveedemk2}
\partial_\kappa m_\kappa^2=-\frac{1}{2}Ng_\kappa(0)\int\frac{d^dq}{(2\pi)^d}\partial_\kappa R_\kappa(q)G^2(\kappa;q) ,
\eeq
and using eq.~(\ref{B3}), it is easy  to check that this equation is equivalent to the gap equation,  eq.~(\ref{gap}).

Next, we observe that in the large $N$ limit, a single channel effectively contributes in eq.~(\ref{4point}) for the $4$-point function;  one can then
use the following identity in this limit:
\begin{eqnarray}\label{B7}
&&\Gamma^{(4)}_{12ij}(\kappa;p_1,p_2,q,\!-\!q\!-\!p_1\!-\!p_2)\Gamma^{(4)}_{34ij}(\kappa;p_3,p_4,\!-\!q,q\!-\!p_3\!-
\!p_4)
\nonumber\\ &&\qquad\qquad=Ng^2_\kappa(p_1+p_2)\delta_{12}\delta_{34},
\end{eqnarray}
together with eq.~(\ref{gamma6N}) for $\Gamma^{(6)}$, and one obtains:
\beq\label{B8}
\kappa\partial_\kappa g_\kappa(p)=Ng_\kappa^2(p) J^{(3)}_d(\kappa;p)-\frac{N}{2}h_\kappa(p)I^{(2)}_d(\kappa),
\eeq
where the function $I^{(2)}_d(\kappa)$ is that defined  in
eq.~(\ref{Ink}), with  $n=2$ . The function $J_d^{(3)}(\kappa;p)$
is obtained from the general definition
\beq
\label{Jnk} J_d^{(n)}(\kappa;p)\equiv\int \frac{d^d q}{(2\pi)^d}
\kappa\partial_\kappa
R_\kappa(q)G^{n-1}(\kappa;q)G(\kappa;p+q).
\eeq
 Note that
$J_d^{(n)}(\kappa;p=0)=I_d^{(n)}(\kappa)$. 

 At this point we remark  that the flow equation for $g_\kappa(p)$
can also be obtained directly from the explicit expression
(\ref{B3}), in the form:
\beq \label{deriveedegk}
\partial_\kappa g_\kappa(p)=-\frac{N}{2}g^2_\kappa(p)\partial_\kappa\int\frac{d^dq}{(2\pi)^d}G(\kappa;q)G(\kappa;q+p).
\eeq
 It is then
straightforward to verify, using eqs.~(\ref{deriveedemk2}) and
(\ref{Hkappa}) that eqs.~(\ref{B8}) and (\ref{deriveedegk}) are
indeed equivalent. The first term
in eq.~(\ref{B8}) comes from the derivative of the cut-off
function in the propagators in eq.~(\ref{deriveedegk}), while the
second term, which involves the 6-point vertex, comes from  the
derivative of the running mass in the propagators.

Note that eqs.~(\ref{deriveedemk2})  for $m_\kappa$ and (\ref{B8}) for $g_\kappa(p=0)$ become identical respectively to eqs.~(\ref{preMk}) and (\ref{preFk}) of the LPA in the large $N$ limit, a well know property \cite{D'Attanasio97}.

Let us now analyze 
characteristic features of the function $g_\kappa(p)$. For simplicity we specialize for the  rest of this subsection to $d=3$. Furthermore, for the purpose of the present, qualitative, discussion, one may assume
$m_\kappa=0$. This allows us to obtain easily
$g_\kappa(p)$ from eq.~(\ref{B3}) in the two limiting cases $p=0$  and $\kappa=0$. In
the first case, we have
\beq\label{B3c}
g_\kappa(0)=\frac{u}{3}\frac{1}{1+\frac{uN}{9\pi^2}\frac{1}{\kappa}}.
\eeq
This is identical to eq.~(\ref{eq:hatg}), with here $\hat
g^*= 1/(2N)$ and $\kappa_c=Nu/9\pi^2$ .  In the
other case, we have
\begin{equation}\label{B3b}
g_{\kappa=0}(p)=\frac{u}{3}\frac{1}{1+\frac{uN}{48}\frac{1}{p}}=
\frac{u}{3}\frac{p}{p+p_c},
\end{equation}
with $p_c\equiv {uN}/{48}$.

\begin{figure}[t!]
\begin{center}
\includegraphics*[width=10cm,angle=0]{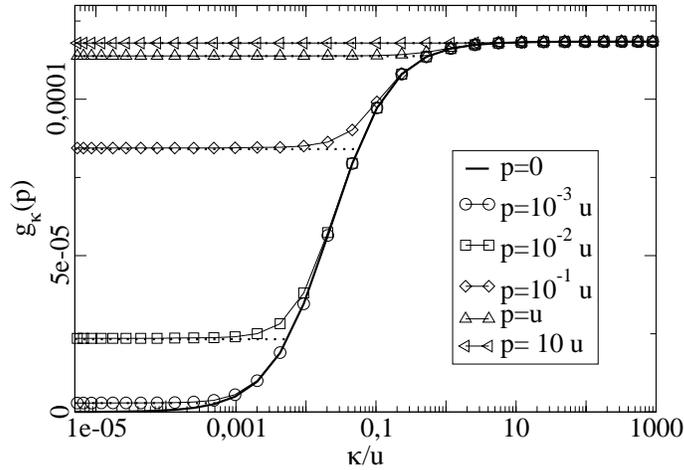}
\end{center}
\caption{\label{fin} The function $g_\kappa(p)$ (in units of $\Lambda$) obtained from a
complete numerical solution of eqs.~(\ref{deriveedemk2}) and
(\ref{B8}), as a function of $\kappa/u$ (in a logarithmic
scale) for five values of $p$: from bottom to top, $p/u=0.001,0.01,0.1,1$ and $10$. The envelope
corresponds to $p=0$. This figure illustrates the decoupling of
modes: for each value of $p$, the flow stops when $\kappa\simle
\alpha p$. The various horizontal asymptotes (dotted llines) correspond to the
single value $\alpha=0.54$. From Ref.~\cite{Blaizot:2005wd}.}
\end{figure}

One sees on eqs.~(\ref{B3c}) and (\ref{B3b}) that the dependence
on $p$ of $g_{\kappa=0}(p)$ is quite similar to the dependence on
$\kappa$ of $g_\kappa(p=0)$. In particular both quantities vanish
linearly as $\kappa\to 0$ or $p\to 0$, respectively. The result of
the complete (numerical) calculation, including the effect of the running  mass ( i.e., solving the gap equation (\ref{gap}) and calculating $g_\kappa(p)$ from eq.~(\ref{B3})),  can in
fact be quite well represented (to within a few percents) for arbitrary $p$ and $\kappa$ by the
following approximate formula \beq\label{gapprox}
g_\kappa(p)\approx \frac{u}{3} \frac{X}{1+X}\qquad X\equiv
\frac{\kappa}{\kappa_c}+\frac{p}{p_c}.
\eeq
This simple expression
shows that $p$, when it is non vanishing, plays the same role as
$\kappa$ as an infrared regulator. In particular, at fixed $p$,
the flow  of $g_\kappa(p)$ stops when $X$ becomes independent of $\kappa$, i.e., when $\kappa\simle p(\kappa_c/p_c)$, with $\kappa_c/p_c=16/3\pi^2\approx 0.54$. This
important property of decoupling of the short wavelength modes is
illustrated in Fig.~\ref{fin}. As shown by this figure, and also
by the expression (\ref{gapprox}), the momentum dependence of the
4-point function can be obtained from its cut-off dependence at zero momentum. In fact Fig.~\ref{fin}
suggests that,  to a very good approximation, there exists a
parameter $\alpha$ such that $g(\kappa;p)\approx g(\kappa;0)$ when
$\kappa>\alpha p$, and $g(\kappa;p)\approx g(\kappa=\alpha p;0)$ when
$\kappa<\alpha p$. From the discussion above, one expects $\alpha\approx \kappa_c/p_c=16/3\pi^2\approx 0.54$, which is indeed in agreement with the analysis in Fig.~\ref{fin}.

\begin{figure}[h!]
\includegraphics[width=10cm,angle=0]{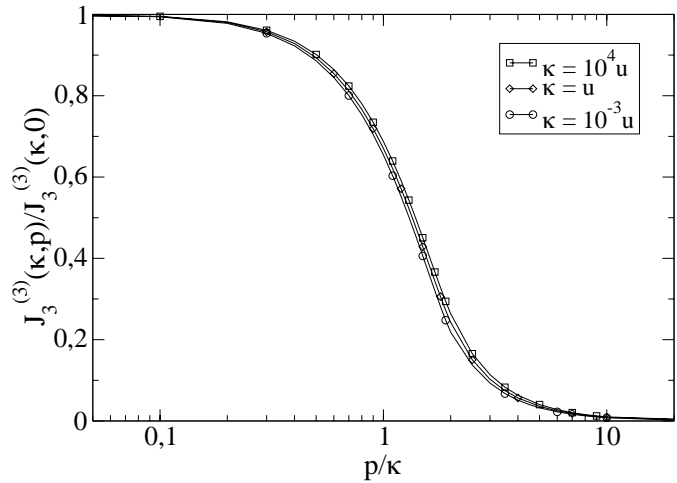}
\caption{\label{fig:theta} The function
$J_3^{(3)}(\kappa;p)/I_3^{(3)}(\kappa)$ as a function of $p/\kappa$
(in a logarithmic scale), for different values of $\kappa$:
$\kappa=10^{-3} u$ (circles), $\kappa=u$ (diamonds) and $\kappa=10^4
u$ (squares). From Ref.~\cite{Blaizot:2005wd}.}
\end{figure}

The decoupling of modes can also be visualized in Fig.~\ref{fig:theta}. The
$p$-dependence of $J_d^{(n)}(\kappa;p)$ (eq.~(\ref{Jnk})) shown in this figure is relatively simple: when
$p\ll \kappa$, $J_d^{(n)}(\kappa;p)\simeq
I_d^{(n)}(\kappa)$; when $p\gg \kappa$, $J_d^{(n)}(\kappa;p)$
vanishes as $1/p^2$. On a logarithmic scale the transition between
these two regimes occurs rapidly at momentum $p\sim \kappa$, as
illustrated on Fig.~\ref{fig:theta}. A similar analysis can be made for the contribtution of the 6-point function in eq.~(\ref{B8}), and one can show that the ratio of the $p$-dependence of the second term in the right hand side (the one proportional to $h_\kappa$) is, whenever it is significant, proportional to that of the first term, the proportionality coefficient being a function of $\kappa$ only. 

All this suggests that one can rewrite
eq.~(\ref{B8}) for $g_\kappa(p)$ as follows: \beq\label{B8b}
\partial_\kappa g_\kappa(p)\approx Ng_\kappa^2(p) \Theta(1-\frac{\alpha^2 p^2}{\kappa^2})I^{(3)}_d(\kappa)(1-F_\kappa),
\eeq where $\alpha$ is a parameter of order unity. The $\Theta$-function ensures that the flow
exists only when $\kappa>\alpha p$, and stops for smaller values
of $\kappa$.

These approximations, together with another one that we shall discuss more thoroughly in the next section have been used in order to construct approximate flow equations in Refs.~\cite{Blaizot:2005wd,Blaizot:2006vr}.

\subsection{Beyond  the derivative expansion}
The arguments on which the derivative expansion is based can be generalized in order to set up a much more powerful approximation scheme \cite{Blaizot:2005xy} that we now briefly present. This scheme allows us to obtain the momentum dependence of the $n$-point function with a single approximation. 
 
  We observe that:
 i)  the momentum $q$ circulating in the loop integral of a flow equation
   is limited by $\kappa$; ii)  the
   smoothness of the $n$-point functions allows us to make an expansion in powers of $q^2/\kappa^2$, independently of the value  of the external momenta $p$. Now, a typical $n$-point function entering a flow equation is of the from $\Gamma^{(n)}_\kappa(p_1,p_2,...,p_{n-1}+q,p_n-q;\phi)$, where $q$ is the loop momentum. The proposed approximation scheme, \emph{in its leading order},  consists in neglecting the  $q$-dependence of such vertex functions:
\begin{equation}\label{approx}
\Gamma^{(n)}_\kappa(p_1,p_2,...,p_{n-1}+q,p_n-q;\phi)\sim
\Gamma^{(n)}_\kappa(p_1,p_2,...,p_{n-1},p_n;\phi).
\end{equation}
Note that this approximation is a priori  well justified.
Indeed, when all the external momenta vanish $p_i=0$, eq.~(\ref{approx})
is the basis of the LPA  which, as stated above, is a good
approximation. When the external momenta $p_i$ begin to grow, the
approximation in eq.~(\ref{approx}) becomes better and better,  and it is
trivial when all momenta are much larger than $\kappa$. With this
approximation,  eq.~(\ref{eq:dGamma2})  becomes (for simplicity we set here $N=1$):
\begin{eqnarray}
\label{gamma2app}
\partial_\kappa\Gamma_\kappa^{(2)}(p,-p;\phi)&=&\int
\frac{d^dq}{(2\pi)^d}\partial_\kappa R_k(q^2)\left\{G_\kappa(q^2;\phi)\Gamma_\kappa^{(3)}(p,0,-p;\phi)\right. \nonumber \\
&&\times G_\kappa((q+p)^2;\phi)\Gamma_\kappa^{(3)}(-p,p,0;\phi)G_\kappa(q^2;\phi) \nonumber \\
&&\left.-\frac{1}{2}G_\kappa(q^2;\phi)\Gamma_\kappa^{(4)}(p,-p,0,0;\phi)G_\kappa(q^2;\phi)\right\}.
\end{eqnarray}

Now comes the second ingredient of the approximation scheme, which
exploits the advantage of working with a non vanishing background
field:  vertices evaluated at  zero external momenta can be
obtained  as derivatives of vertex functions with a smaller number
of legs:
\begin{equation}
\label{faireq=0}
\Gamma_\kappa^{(n+1)}(p_1,p_2,...,p_n,0;\phi)=\frac{\partial
\Gamma_\kappa^{(n)}(p_1,p_2,...p_n;\phi)} {\partial \phi}.
\end{equation}
 By exploiting eq.~(\ref{faireq=0}), one easily
transforms eq.~(\ref{gamma2app}) into a {\it closed equation} (recall
that $G_\kappa$ and $\Gamma_\kappa^{(2)}$ are related by
eq.~(\ref{Ggamma2})):
\begin{eqnarray}\label{2pointclosed}
&&\partial_\kappa\Gamma_\kappa^{(2)}(p^2;\phi)=\int
\frac{d^dq}{(2\pi)^d}\,\partial_\kappa R_\kappa(q^2) \; G^2_\kappa(q^2;\phi)  \nonumber \\
&&
\times \left\{ \left( \frac{\partial \Gamma_\kappa^{(2)}(p,-p;\phi)}
{\partial \phi} \right)^2 G_\kappa ((p+q)^2;\phi)
\; - \; \frac{1}{2}\frac{\partial^2 \Gamma_\kappa^{(2)}(p,-p;\phi)}
{\partial \phi^2}\right\}.
\end{eqnarray}

Note the similarity of this equation with  
eq.~(\ref{eq-pot}): both are closed equations,
the vertices appearing in the r.h.s.   being expressed
as derivatives of the function in the l.h.s.. 

The approximation scheme presented here is similar to that used in \cite{Blaizot:2005wd,Blaizot:2006vr}. There also the momentum dependence of the vertices was neglected in the leading order. However further approximations were needed in order to close the hierarchy. The progress realized here is to bypass these extra approximations by working in a constant background field.

The resulting equations can be solved with a numerical effort comparable to that involved in solving the equations of the derivative expansion. The preliminary results obtained so far are encouraging \cite{Blaizot:2006ak}.

\subsection{Calculation of $\Delta\langle \varphi^2\rangle$}

We come now to the conclusion of these lectures, where results concerning the numerical value of $c$ obtained with the NPRG will be presented. 

The approximation developed in \cite{Blaizot:2004qa,Blaizot:2005wd,Blaizot:2006vr} is based on an iteration scheme, where one starts by building   approximate equations for the $n$-point functions, that are then solved exactly. To give a flavor of this method, consider more specifically the equation (\ref{4point}) for the 4-point function. An approximate flow equation is obtained with the following three approximations: 

 Our first approximation is that discussed in sect.~4.4: we 
ignore the $q$ dependence of the vertices in the flow equation,
i.e., we set $q=0$ in the vertices $\Gamma^{(4)}$ and
$\Gamma^{(6)}$ and factor them out of the integral in the r.h.s.
of eq.~(\ref{4point}). 
The second approximation  concerns the propagators in the flow equation, for which we make the replacements:
\beq\label{A2}
G(p+q)\longrightarrow G_{LPA'}(q)\,\Theta\left(1-\frac{\alpha^2 p^2}{\kappa^2}\right)
\eeq
where $\alpha$ is an adjustable parameter. A motivation for this approximation may be obtained from the analysis of the $n$-point function presented in sect. 4.3 (see eq.~(\ref{B8b}) and Fig.~\ref{fig:theta}).  
 This approximation introduces a dependence of the  
results on the value of $\alpha$.    The third approximation  concerns the function $\Gamma^{(6)}$ for which we
use an ansatz inspired by the  expressions of the various $n$-point functions in the large $N$ limit (see again the discussion at the end of sect. 4.3): one assumes that the contribution of $\Gamma^{(6)}$ to the
r.h.s. of eq.~(\ref{4point}) is proportional to
 that of  the other terms, the proportionality coefficient $F_\kappa$ being only a function
of $\kappa$. The same proportionality also holds in
the LPA regime, which allows us to use the LPA to determine
$F_\kappa$.
 
 These three approximations  result then in  the following approximate equation for $\Gamma^{(4)}$:
\small\beq
\label{gamma40new}
&&\hspace{-0.6cm}\partial_\kappa\Gamma^{(4)}_{ijkl}(\kappa;p_1,p_2,p_3,p_4)=I^{(3)}_\kappa(0)\,
(1-F_\kappa)
  \nonumber \\
&&\hspace{.1cm}\times\left\{ \Theta\!\!\left(\kappa^2-\alpha^2
{(p_1+p_2)^2})\right)
\Gamma^{(4)}_{ijmn}(\kappa;p_1,p_2,0,-p_1-p_2)\right. \nonumber\\
&&\hspace{1cm}\left.\times\Gamma^{(4)}_{klnm}(\kappa;p_3,p_4,-p_3-p_4,0)
+{\rm permutations}
\right\}.
\eeq\normalsize
This equation  can be solved analytically in
terms of the solution of the LPA (actually a refined version of it). This is done by steps,
starting form the momentum domain
$\alpha^2(p_1+p_2)^2,\alpha^2(p_1+p_3)^2,\alpha^2(p_1+p_4)^2\le
\kappa^2$, where it can be verified that the solution is  that of the LPA itself. The solution of this equation is then inserted in eq.~(\ref{eq:dGamma2}) for the 2-point function and leads to the leading order determination of $\Sigma(p)$. The scheme is improved through an iteration procedure. At next-to-leading order, one improves $\Gamma^{(4)}$ by establishing an approximate equation for $\Gamma^{(6)}$. The solution of this equation is then used in eq.~(\ref{4point}) for  $\Gamma^{(4)}$, and the resulting $\Gamma^{(4)}$ is used in turn in eq.~(\ref{eq:dGamma2}) to get $\Sigma$ at next-to-leading order. 
 \begin{table}
\begin{tabular}{|c|c|c|c|c|c|c|c|}
\hline $c$ & $N=1$ & $N=2$ & $N=3$ & $N=4$ & $N=10$ & $N=40$ &
$N=\infty$
\\ \hline
lattice \cite{latt2} & & $1.32 \pm 0.02$ & & & & & \\
lattice \cite{latt1} & & $1.29 \pm 0.05$& & & & & \\
lattice \cite{latt3} & $1.09\pm 0.09$ & & & $1.60\pm 0.10$ & & & \\
7-loops \cite{Kastening:2003iu} & $1.07\pm 0.10$ & $ 1.27 \pm 0.10$ & $1.43\pm 0.11$ & $1.54\pm 0.11$ & & & \\
large $N$ \cite{BigN} & & & & & & &$c=2.33$ \\
NPRG & 1.11 & 1.30  & 1.45 & 1.57 & 1.91 & 2.12 & \\
\hline
\end{tabular}
\caption{Summary of available results for the coefficient $c$. The
last line contains the results obtained in \cite{Blaizot:2006vr}. }
\end{table}

\begin{figure}[t!]
\begin{center}
\includegraphics*[width=12 cm,angle=0]{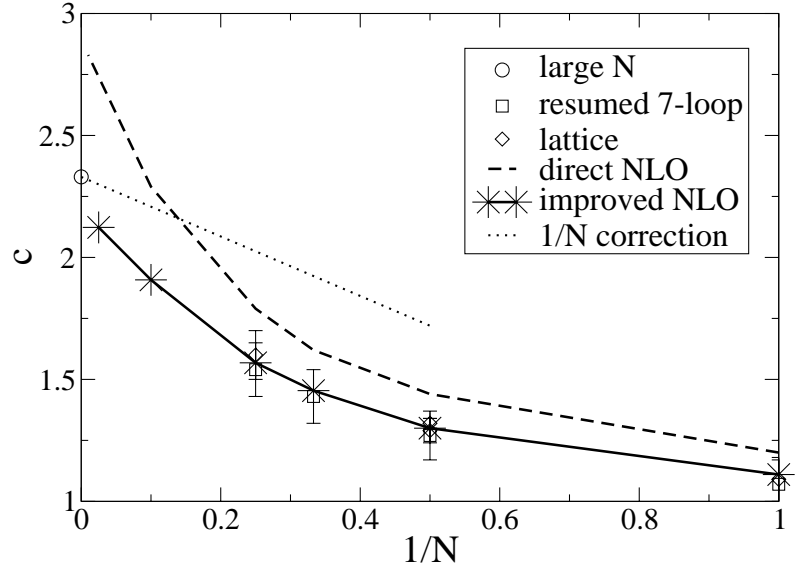}
\end{center}
\caption{\label{coefc_N} The coefficient $c$ (obtained with the
fastest apparent convergence procedure) as a
function of $1/N$. Our NLO results, are compared with
results obtained with other methods: lattice \cite{latt2,latt1,latt3}
(diamonds) and 7-loops perturbation theory \cite{Kastening:2003iu}
(squares), all of them with their corresponding error bars, together
with the $N\to\infty$ result \cite{BigN} (circle), and the extrapolation following the $1/N$ correction calculated in Ref.~\cite{Arnold:2000ef}. From Ref.~\cite{Blaizot:2006vr}.}
\end{figure}

 The results obtained with this method for the value of $c$ are reported in Table 1 and Fig.~\ref{coefc_N} for various values of $N$
for which results have been obtained with other techniques, either
the lattice technique \cite{latt2,latt1,latt3}, or variationally
improved 7-loops perturtbative calculations \cite{Kastening:2003iu} (other optimized perturbative calculations have also been recently
performed, and are in agreement with those quoted here; see
\cite{souza,Kneur04}).  The results reported in Table 1 have been obtained with an improved next-to-leading order calculation where, among other things, the parameter $\alpha$ of eq.~(\ref{A2}) has been fixed by a principle of minimum sensitivity (see \cite{Blaizot:2006vr} for details). A recent (approximate) calculation along the lines described in sect.~4.4 yields, for $N=1$, the value $c=1.2$ \cite{Blaizot:2006ak}. Let me also mention the result $c=1.23$ obtained, for $N=2$ in Ref.~\cite{Ledowski03}; however, as discussed in \cite{Blaizot:2006vr} it is difficult to gauge the quality of the approximation made in \cite{Ledowski03}.

  For $N\le 4$, where we can compare
with other results, the values of $c$ obtained with the present
improved NLO calculation are in excellent agreement with those
obtained from lattice and ``7-loops'' calculations. What happens at large values of $N$ deserves a special discussion.
As seen in Fig.~\ref{coefc_N} the curve showing the improved leading
order results extrapolates  when $N\to \infty$ to a value that is
about $4\%$ below the known exact result \cite{BigN}. A direct
calculation at very large values of $N$ is difficult in the present
approach for numerical reasons: since the coefficient $c$ represents
in effect an order $1/N$ correction (see \cite{BigN}), it is
necessary to insure the cancellation of  the large, order $N$,
contributions to the self-energy, in order to extract the value of
$c$. This is numerically demanding when $N\simge 100$.
Fig.~\ref{coefc_N} also reveals an intriguing feature: there seems
to be no natural way to reconcile the present results, and for this
matter the results from lattice calculations or 7-loop calculations,
with the calculation of the $1/N$ correction presented in
Ref.~\cite{Arnold:2000ef}: the dependence in $1/N$ of our results,
be they obtained from the direct NLO or the improved NLO, appear to be
incompatible with the slope predicted in Ref.~\cite{Arnold:2000ef} for the $1/N$ expansion.

Acknowledgements. Most of the material of these lectures is drawn from work done in a most enjoyable collaboration with several  people during these last years: G. Baym, F. Lalo\"{e}, M. Holzmann,  R. Mendez-Galain, D. Vautherin, N. Wschebor, J. Zinn-Justin. I would also like to express my gratitude  to Achim Schwenk for insisting on having these lecture notes....and putting such a high cut-off on his patience as he waited for them.
%
% BibTeX users please use
%\bibliographystyle{unsrt}
%\bibliography{Bose_Einstein,jpb,functionalRG}

\begin{thebibliography}{0}
%\input{referenc}

%%%%%%%%%%%%%%%%%%%%%%%%%%%%%%%%%%%%%%%%%%%%%%%%%%%%%%%%%%%%%%%%%%%%%%  }

%%%%%%%%%%%%%%%%%%%%%%%%%%%%%%%%%%%%%%%%%%%%%%%%%%%%%%%%%%%%%%%%%%%%%%

\bibitem{Pitaevskii03}
L. Pitaevskii and S. Stringari, {\it Bose-Eisntein condensation}, Oxford University Press, UK, 2003.

\bibitem{Pethick02} 
C.J. Pethick and H. Smith, {\it Bose-Eisntein condensation in dilute gases}, Cambridge University Press, UK, 2003.

  \bibitem{RMP} F.\ Dalfovo, S. Giorgini, S.\ Stringari, and L.\
Pitaevskii, Rev.\ Mod.\
Phys.\ {\bf 71}, 463 (1999).

\bibitem{Leggett:2001}
A.~J. Legget.
\newblock Bose-einstein condensation in the alkali gases: Some fundamental
  concepts.
\newblock {\em Rev. Mod. Phys.}, 73:307, 2001.

%\cite{Andersen:2003qj}
\bibitem{Andersen:2003qj}
  J.~O.~Andersen,
  %``Theory of the weakly interacting Bose gas,''
  Rev.\ Mod.\ Phys.\  {\bf 76} (2004) 599
  %[arXiv:cond-mat/0305138].
  %%CITATION = RMPHA,76,599;%%

      \bibitem{3/2club} G.\ Baym, J-P.\ Blaizot, M.\ Holzmann, F.\ 
Lalo\"{e} and
D.\ Vautherin, Phys.\ Rev.\ Lett.  {\bf 83}, 1703 (1999).

       \bibitem{BigN} G.\ Baym, J-P.\ Blaizot and J. Zinn-Justin, Europhys.\
Lett.  {\bf 49}, 150 (2000).


\bibitem{bigbec}
G.\ Baym, J-P.\ Blaizot, M.\ Holzmann, F.\ 
Lalo\"{e} and
D.\ Vautherin, Eur. Phys. J. {\bf B 24} (2001) 107.

       \bibitem{NEW} M. Holzmann, G. Baym, J.-P. Blaizot, and F. Lalo\"e,
Phys. Rev. Lett. {\bf 87}, 120403 (2001).

\bibitem{Fuchs04}  M. Holzmann, J.N. Fuchs, G. Baym, J.-P. Blaizot, and F. Lalo\"e,
Comptes Rendus Physique {\bf 5} (2004)21.
 
  \bibitem{Blaizot:2004qa}
  J.~P.~Blaizot, R.~Mendez Galain and N.~Wschebor,
  Europhys. Lett., {\bf 72 (5)}, 705-711 (2005).
  
 %\cite{Blaizot:2005wd}
\bibitem{Blaizot:2005wd}
  J.~P.~Blaizot, R.~Mendez-Galain and N.~Wschebor,
  %``Non perturbative renormalisation group and momentum dependence of n-point
  %functions. I,''
  Phys. Rev. E74 (2006) 051116.
  % [arXiv:hep-th/0512317].
  %%CITATION = HEP-TH 0512317;%%


%\cite{Blaizot:2006vr}
\bibitem{Blaizot:2006vr}
  J.~P.~Blaizot, R.~Mendez-Galain and N.~Wschebor,
  %``Non perturbative renormalization group and momentum dependence of n-point
  %functions. II,''
  Phys. Rev. E74 (2006) 051117.
  % [arXiv:hep-th/0603163].
  %%CITATION = HEP-TH 0603163;%%
  
  %\cite{Blaizot:2005xy}
\bibitem{Blaizot:2005xy}
  J.~P.~Blaizot, R.~Mendez Galain and N.~Wschebor,
  %``A new method to solve the non perturbative renormalization group
  %equations,''
  Phys.\ Lett.\ B {\bf 632}, 571 (2006).
  %[arXiv:hep-th/0503103].
  %%CITATION = HEP-TH 0503103;%%
 

  %\cite{Blaizot:2006ak}
\bibitem{Blaizot:2006ak}
  J.~P.~Blaizot, R.~Mendez-Galain and N.~Wschebor,
  %``Non-perturbative renormalization group calculation of the scalar
  %self-energy,''
  arXiv:hep-th/0605252.
  %%CITATION = HEP-TH/0605252;%%
  
  
  \bibitem{Ziff77}
R.M. Ziff, G.E. Uhlenbeck and M. Kac, Phys. Rep. {\bf 32} (1977) 169.

\bibitem{PNAS}  M. Holzmann, G. Baym, J.-P. Blaizot, F. Lalo{\"e}, 
%arXiv:cond-mat/0508131
PNAS 2007 104: 1476-1481.


\bibitem{Giorgini96}
S. Giorgini, L. Pitaevskii and S. Stringari, Phys. Rev. {\bf A54} (1996) 4633.

%\cite{Schafer:2006yf}
\bibitem{Schafer:2006yf}
  T.~Schafer,
  %``Effective theories of dense and very dense matter,''
  arXiv:nucl-th/0609075.
  %%CITATION = NUCL-TH/0609075;%%

%\cite{Kaplan:2005es}
\bibitem{Kaplan:2005es}
  D.~B.~Kaplan,
  %``Five lectures on effective field theory,''
  arXiv:nucl-th/0510023.
  %%CITATION = NUCL-TH/0510023;%%

%\cite{Braaten:1996rq}
\bibitem{Braaten:1996rq}
  E.~Braaten and A.~Nieto,
  %``Renormalization effects in a dilute Bose gas,''
  arXiv:hep-th/9609047.
  %%CITATION = HEP-TH/9609047;%

%\cite{Braaten:1997ky}
\bibitem{Braaten:1997ky}
  E.~Braaten and A.~Nieto,
  %``Quantum corrections to the ground state of a trapped Bose-Einstein
  %condensate,''
  arXiv:cond-mat/9707199.
  %%CITATION = COND-MAT/9707199;%%

%\cite{Braaten:2000eh}
\bibitem{Braaten:2000eh}
  E.~Braaten, H.~W.~Hammer and S.~Hermans,
  %``Nonuniversal Effects in the Homogeneous Bose Gas,''
  Phys.\ Rev.\  A {\bf 63} (2001) 063609.
 % [arXiv:cond-mat/0012043].
  %%CITATION = PHRVA,A63,063609;%%

       \bibitem{Toyoda} T.\ Toyoda, Annals of Physics N.Y.  {\bf 141}, 154
(1982).

    \bibitem{FW} A.L.\ Fetter and J.D.\ Walecka, {\em Quantum theory of
many-particle systems}, McGraw-Hill (1971),  28.

\bibitem{BG} 
G. Baym and G. Grinstein, Phys.\ Rev.\ {\bf D15}, 2897
(1977).

\bibitem{KB} L.P.\ Kadanoff and G. Baym, {\em Quantum 
statistical
mechanics}, Benjamin (1962).

 
\bibitem{self-consistent} A. A. Abrikosov, L.P.  Gorkov, and 
I.E.\
Dzyaloshinski, {\em Methods of quantum field theory in 
statistical physics},
Prentice Hall (1963).


      \bibitem{BR} J.P.\ Blaizot and G.\ Ripka, 
{\em Quantum theory of finite
systems}, MIT\ Press (1986).

 
      \bibitem{PPbook} 
A.Z.\ Patashinskii and V.L.\ Pokrovskii, {\em Fluctuation
theory of 
phase transitions}, Pergamon Press (1979).

\bibitem{Gins80}
P. Ginsparg, \NPB{170}{1980}{388}.

\bibitem{Appel81}
T. Appelquist and R.D. Pisarski, \PRD{23}{1981}{2305}.

\bibitem{Nadkarni83}
S. Nadkarni, \PRD{27}{1983}{917}; \PRD{38}{1988}{3287}.

\bibitem{arnold01}
  P.~Arnold and B.~Tomasik, Phys. Rev. {\bf A64} (2001) 053609.
  
  %\cite{Moshe:2003xn}
\bibitem{Moshe:2003xn}
  M.~Moshe and J.~Zinn-Justin,
  %``Quantum field theory in the large N limit: A review,''
  Phys.\ Rept.\  {\bf 385}, 69 (2003)
 % [arXiv:hep-th/0306133].
  %%CITATION = HEP-TH 0306133;%%
  
  %\cite{Arnold:2000ef}
\bibitem{Arnold:2000ef}
  P.~Arnold and B.~Tomasik,
  %``T_c for dilute Bose gases: beyond leading order in 1/N,''
  Phys.\ Rev.\  A {\bf 62} (2000) 063604.
  %[arXiv:cond-mat/0005197].
  %%CITATION = PHRVA,A62,063604;%%
  
  %\cite{Zinn-Justin:2002ru}
\bibitem{Zinn-Justin:2002ru}
  J.~Zinn-Justin,
  \emph{Quantum field theory and critical phenomena},
  Int.\ Ser.\ Monogr.\ Phys.\  {\bf 113},  1 (2002).
  %%CITATION = IMPHA,113,1;%%
  
\bibitem{latt2} P. Arnold and G. Moore,
Phys. Rev. Lett. {\bf 87}, 120401 (2001); Phys. Rev. {\bf E64},
066113 (2001).
 
\bibitem{latt1} V.A.  Kashurnikov, N.~V.  Prokof'ev, and B.~V.
Svistunov, Phys. Rev. Lett. {\bf 87}, 160601 (2001); Phys. Rev.
Lett. {\bf 87}, 120402 (2001).
  
 
\bibitem{Kneur04}
J.-L. Kneur, A. Neveu, M. B. Pinto, Phys. Rev. {\bf A69}, 053624
(2004).

\bibitem{souza} F. de Souza Cruz, M.B. Pinto, and R.O. Ramos,
Phys. Rev. {\bf B64}, 014515 (2001); Phys.\ Rev.\ A {\bf 65}, 053613
(2002).
  
   \bibitem{Kastening:2003iu}
B.~Kastening, Phys.\ Rev.\ A {\bf 68}, 061601 (R) (2003); Phys.\
Rev.\ A {\bf 69}, 043613 (2004).
%%CITATION = COND-MAT 0309060;%%

%\cite{Arnold:2001nn}
\bibitem{Arnold:2001nn}
  P.~Arnold, G.~D.~Moore and B.~Tomasik,
  %``T_c for homogeneous dilute Bose gases: a second-order result,''
  Phys.\ Rev.\  A {\bf 65} (2002) 013606
 % [arXiv:cond-mat/0107124].
  %%CITATION = PHRVA,A65,013606;%%










\bibitem{Wetterich93}  C.Wetterich, Phys. Lett., {\bf B301}, 90 (1993).

\bibitem{Ellwanger93}  U.Ellwanger, Z.Phys., {\bf C58}, 619 (1993).

%\cite{Tetradis:1993ts}
\bibitem{Tetradis94}
  N.~Tetradis and C.~Wetterich,
  %``Critical exponents from effective average action,''
  Nucl.\ Phys.\ B {\bf 422},  541 (1994).
  %[arXiv:hep-ph/9308214].
  %%CITATION = HEP-PH 9308214;%%

\bibitem{Morris94}  T.R.Morris, Int. J. Mod. Phys., {\bf A9}, 2411 (1994).

\bibitem{Morris94c}  T.R.Morris, Phys. Lett. {\bf B329}, 241 (1994).


\bibitem{Bagnuls:2000ae}
C.~Bagnuls and C.~Bervillier,
%``Exact renormalization group equations: An introductory review,''
Phys.\ Rept.\  {\bf 348}, 91 (2001).

\bibitem{Berges02}
J. Berges, N. Tetradis and C. Wetterich,
   Phys. Rept. {\bf 363}, 223--386 (2002).

\bibitem{Canet04} B.~Delamotte, D.~Mouhanna, M.~Tissier, Phys. Rev. {\bf B69}, 134413 (2004);  L. Canet and B. Delamotte, cond-matt/0412205.





%\cite{Gies:2006wv}
\bibitem{Gies:2006wv}
  H.~Gies,
  %``Introduction to the functional RG and applications to gauge theories,''
  arXiv:hep-ph/0611146.
  %%CITATION = HEP-PH/0611146;%%

       \bibitem{Stoof2} M.\ Bijlsma and H.T.C.\ Stoof, Phys.\ Rev.\ {\bf A54},
5085 (1996).
 
 
 %\cite{Andersen:1998bc}
\bibitem{Andersen:1998bc}
  J.~O.~Andersen and M.~Strickland,
  %``Critical behavior of a homogeneous Bose gas at finite temperature,''
  arXiv:cond-mat/9808346.
  %%CITATION = COND-MAT/9808346;%%
 
 %\cite{Andersen:1999dy}
\bibitem{Andersen:1999dy}
  J.~O.~Andersen and M.~Strickland,
  %``Application of Renormalization Group Techniques to a Homogeneous Bose Gas
  %at Finite Temperature,''
  Phys.\ Rev.\  A {\bf 60} (1999) 1442.
 % [arXiv:cond-mat/9811096].
  %%CITATION = PHRVA,A60,1442;%%

\bibitem{Ball95} R.D.Ball, P.E.Haagensen, J.I.Latorre and E. Moreno,
Phys. Lett., {\bf B347}, 80 (1995).

\bibitem{Comellas98} J.Comellas, Nucl. Phys., {\bf B509}, 662 (1998).

\bibitem{Litim}
D.Litim, Phys. Lett. {\bf B486}, 92 (2000); Phys. Rev. {\bf D64},
105007 (2001);  Nucl. Phys. {\bf B631}, 128 (2002);
Int.J.Mod.Phys. {\bf A16}, 2081 (2001).

\bibitem{Canet02} L.Canet, B.Delamotte, D.Mouhanna and J.Vidal, Phys.
Rev. {\bf D67}, 065004 (2003).


\bibitem{Ellwanger94} U. Ellwanger and C. Wetterich, Nucl. Phys. {\bf B423}, 137(1994).

\bibitem{Tetradis95}
N.~Tetradis and D.~F. Litim,
%\newblock Analytical solutions of exact renormalization group equations.
Nucl. Phys.  {\bf B464},  492--511 (1996).

\bibitem{D'Attanasio97}
M. D'Attanasio and T.~R. Morris,
%\newblock Large {N} and the renormalization group.
Phys. Lett. {\bf B409},  363--370 (1997).

\bibitem{latt3} X.~Sun, Phys. Rev. {\bf E67}, 066702 (2003).

\bibitem{Ledowski03}  S.Ledowski, N. Hasselmann and P. Kopietz, Phys. Rev. {\bf A 69}, 061601(R)
(2004); Phys. Rev. {\bf A 70}, 063621 (2004).
   
 
\end{thebibliography}
%
% Non-BibTeX users please follow the syntax
% the syntax of "referenc.tex" for your own citations

%\printindex
\end{document}